\documentclass[%
 reprint,
 amsmath,amssymb,
 aps,
 pre,
]{revtex4-2}

\usepackage{graphicx,lipsum}
\usepackage{dcolumn}
\usepackage{bm}
\usepackage{subfigure}
\usepackage{multirow}
\usepackage{makecell}
\usepackage[section]{placeins}
\usepackage{multirow}
\usepackage[margin=1.5cm]{geometry}
\usepackage[colorlinks=true,linkcolor=blue,citecolor=blue,urlcolor=blue,bookmarks=false,hypertexnames=true]{hyperref}
\usepackage{epsf,epsfig,graphics}
\usepackage{orcidlink}
\usepackage[utf8]{inputenc}
\begin{document}

\preprint{APS/123-QED}

\title{Spectral form factor and power spectrum for trapped interacting rotating bosons: Crossover from integrability to quantum chaos}

\author{Mohd Talib \orcidlink{0000-0001-7261-3584}}
\email{rs.mtalib@jmi.ac.in}
\author{M. A. H. Ahsan \orcidlink{0000-0002-9870-2769}}%
\email{mahsan@jmi.ac.in}
\affiliation{Department of Physics, Jamia Millia Islamia (A Central University), New Delhi, 110025, India.}%

\date{\today}

\begin{abstract}
The emergence of quantum chaos in a system of trapped interacting bosons with externally impressed rotation is studied through spectral form factor (SFF) and power spectrum using exact diagonalization. Two distinct interaction regimes are considered: the moderate, when the interaction energy is small compared to the trap energy and the strong, when the interaction energy is comparable to the trap energy. In the moderate interaction regime, the SFF for the non-rotating case exhibits a dip-plateau structure with absence of linear ramp, indicating integrable behavior, while for the single-vortex state the SFF exhibits a discernible linear ramp consistent with pseudo-integrable behavior. In the strong interaction regime, the non-rotating case exhibits emergence of a linear ramp with small time span in SFF, indicating that the system has moved further towards chaotic regime but continues to be pseudo-integrable. For the single-vortex and the multi-vortex states in strong interaction regime, the span of the linear ramp in SFF increases progressively with rotation, indicating the system has moved into strong chaotic regime consistent with Gaussian orthogonal ensemble. The power spectrum results with exponent lying in the interval $1 \lesssim \alpha \lesssim 2$ are consistent with the findings of SFF. An understanding of the observed crossover from integrable to quantum chaos is presented in terms of the macroscopic occupation of a single-particle quantum state---the Bose-Einstein condensation---and its depletion driven by interaction and rotation.

\end{abstract}

\maketitle

\section{Introduction} In recent years, the study of quantum chaos in many-body systems has developed into a field of intensive research \cite{Guhr, Tianci, Fausto, Lea2020}. This has largely been driven by its relevance to phenomenon such as eigenstate thermalization \cite{Deutsch, Srednicki, Hosur, Liu2014} and scrambling of quantum information in many-body systems \cite{Iyoda2018, Yuan2022}. 
The quantum chaotic behavior has been extensively explored across a broad class of many-body quantum systems, including interacting bosonic and fermionic models \cite{Santos2010, santos, Maier2025,Munoz2006, Deota2013, CHAVDA2025}, spin-$1/2$ systems \cite{santos13}, mesoscopic systems such as quantum dots \cite{Alhassid}, etc. The energy-level based indicators of quantum chaos, including the nearest-neighbour spacing distribution \cite{mehta, stockmann}, the spectral form factor \cite{haake} and the power spectrum \cite{Relano2002}, as well as wavefunction-based diagnostics such as out-of-time-order correlators (OTOCs) \cite{Maldacena2016, Rozenbaum2017}, have emerged as powerful tools to investigate quantum chaos and have attracted considerable attention within the quantum many-body community.

A precise and universally acceptable definition of quantum chaos is yet to be established. The Berry-Tabor \cite{berry} and the Bohigas-Giannoni-Schmit (BGS) \cite{bohigas} conjectures provide a framework to study quantum chaos via random matrix theory (RMT), developed historically by Wigner \cite{wigner55,wigner57,wigner58} to describe the statistical properties of complex nuclear spectra. A paradigmatic framework to investigate quantum chaos in finite fermionic systems is provided by complex nuclei leading to advancing our understanding of chaos in quantum many-body systems \cite{Weidenmullerrmp2009,Mitchell2010, GOMEZ2011pr}.  

The low energy spectra of deformed nuclei are described by the collective model \cite{Mottelson1975, Heyde2010} wherein a nucleus consists of a core of inner closed shells of nucleons and outer valence nucleons that behave analogously to surface molecules in a liquid drop. The motion of valence nucleons acts as a perturbation to induce nonsphericity in the inner core which in turn modifies the quantum states including rotational states of valence nucleons. Nuclear matter exhibits superfluidity due to pairing correlations arising from the attractive nuclear force \cite{Brink2023,Dean2003,Coello2025, Sedrakian2019, Litvinova2020}. Increasing rotation weakens pairing correlations, due to the so called Coriolis-antipairing effect \cite{Dean2010}, bridging collective superfluid (bosons) behavior with single-particle fermionic behavior \cite{OTSUKA1978, Mottelson1960, Hung2011, Quentin2020}.

The late-time behavior is reveal by the spectral form factor (SFF) which captures both short-range and long-range spectral correlations. The SFF exhibits three characteristic time regimes corresponding to the slope, the ramp, and the plateau. The slope corresponds to early-time dynamics, while the ramp characterises level repulsion and serves as a hallmark of quantum chaos. At late times, the SFF saturates, giving rise to plateau, which corresponds to discreteness of the spectrum. There were a series of studies that established the relevance of power spectrum to quantum chaos \cite{Relano2002, Faleiro2004, Relano2008, Gomez2005, Relanopre2006}. The power spectrum provides an additional statistical tool to analyze both short-range and long-range correlations. Integrable systems are associated with $1/f^2$ noise, while chaotic systems exhibit $1/f$ noise. For pseudo-integrable systems, the power-law exhibits $1/f^{\alpha}$ behavior with power exponent lying in the interval $1 < \alpha < 2$.

The SFF has been employed extensively across a wide range of model systems, including Sachdev-Ye-Kitaev model \cite{caceres2022spectral, orman2025quantum}, black hole systems \cite{cotler2017black, Garcia2016, Phil2018semiclassical, Junyu2018}, Aubry-Andr\'e-Harper model \cite{Aamna2021}, etc. It has also been used as a diagnostic of phenomena such as chaos \cite{Amos2018, Pavel2018, Bruno2018, DasPRR2025} and localization in many-body quantum systems \cite{Jan2020, Abhishodh2021, Garratt2021}. The SFF has recently been measured experimentally on superconducting quantum processors to probe quantum chaos in many-body systems \cite{Hang2025}. The utility of the power spectrum has firmly been established in quantum chaos studies across diverse systems including quartic oscillator, kicked top \cite{Santhanam2005} and more recently in molecular resonance studies of erbium isotopes \cite{Mur-Petit2015}.
 
The main objective of this article is to explore crossover from integrability to chaos in Bose-Einstein-condensed harmonically trapped interacting bosons employing the spectral form factor and the power spectrum as diagnostic tools. The strong interparticle interaction and the nucleation of vortices causes depletion of the condensate \cite{ueda2006, Tomadin2011}, enhancing spectral correlations and driving the system towards chaotic behavior. Although the effects of interaction and rotation are well studied in superfluid systems \cite{Mahan2013} and trapped bosons \cite{Ahsan2001, Xiaprl2001, Xiapra2001}, their impact on crossover from integrable to chaotic behavior has remained largely unexplored. Experimentally, in ultracold trapped atomic gases, the interparticle interaction can be tuned via the Feshbach resonance \cite{inouye1998,chin2010} and vortices may be generated using focused laser beam \cite{Madison}.

The remainder of the paper is organized as follows. Sec. \ref{1} introduces a model system exhibiting Bose-Einstein condensation. In Sec. \ref{2}, we discuss the tools employed to study quantum chaos, specifically the spectral form factor and the power spectrum. The numerical results are presented in Sec. \ref{3}, and Sec. \ref{4} concludes the paper with a summary and outlook for future work. 

\section{\label{1} Model system exhibiting Bose-Einstein condensation}
 
A realistic many-body system under certain physically realizable conditions reduces to a two-dimensional system exhibiting rich physical phenomena like fractional quantum Hall effect (FQHE) \cite{Tsui1982}, Berezinskii-Kosterlitz-Thouless (BKT) transition  \cite{Berezinskii1971, Berezinskii1972, Kosterlitz_1973} etc. 

We consider a system of $N$ interacting bosons confined in a cylindrically symmetric harmonic trap. The Hamiltonian for the system in the laboratory frame is given by
\begin{align}
\hat{H} 
=&\sum_{i=1}^{N}\!\left[\frac{1}{2M}\!\left(\frac{\hbar}{i}\nabla_{\perp i}\right)^{2} + \frac{1}{2}M\omega_{\perp}^{2}r_{\perp i}^{2}\right] \nonumber\\ &+ \sum_{i=1}^{N}\!\left[\frac{1}{2M}\!\left(\frac{\hbar}{i}\nabla_{z_{i}}\right)^{2} + \frac{1}{2}M\omega_{z}^{2}z_{i}^{2}\right] \nonumber\\
&+ \frac{1}{2}\frac{4\pi\hbar^{2}a_{sc}}{M}
\left(\frac{1}{\sqrt{2\pi}\sigma}\right)^{3} \nonumber\\
&\times \sum_{i < j}^{N}\! \exp\!\left[-\frac{(r_{\perp i}-r_{\perp j})^{2}+(z_{i}-z_{j})^{2}}{2\sigma^{2}}\right], \label{hlab}
\end{align}
where $a_{sc}$ is the $s$-wave particle-particle scattering length, $M$ being the mass of the boson and $\lambda_{z} \equiv \omega_{z}/\omega_{\perp}$ is the anisotropy parameter, with $\omega_{\perp}$ and $\omega_{z}$ denoting the radial and the axial trapping frequencies, respectively.
\noindent The one-body non-interacting Hamiltonian for motion along the z-direction is given by

\begin{align}\label{a1}
    \hat{h}_{z}=\!\frac{1}{2M}\!\left(\frac{\hbar}{i}\nabla_{z}\right)^{2} 
+ \frac{1}{2}M\omega_{z}^{2}z^{2},
\end{align}
and the corresponding eigensolution for $\hat{h}_{z}u_{n_{z}}(z)=\left(n_{z}+\frac{1}{2} \right)\hbar \omega_{z} u_{n_{z}}(z)$ with $\quad n_{z} = 0, 1, 2, \ldots$, is given by
\begin{equation}
u_{n_{z}}(z) = 
\sqrt{\frac{\alpha_{z}}{\sqrt{\pi}\,2^{n_{z}}n_{z}!}}
\, e^{-\tfrac{1}{2}\alpha_{z}^{2}z^{2}}
H_{n_{z}}(\alpha_{z}z),
\end{equation}
where $\alpha_{z}\equiv\sqrt{\frac{M\omega_{z}}{\hbar}}$ with $\omega_{z}$ being the axial frequency and $H_{n_z}(\alpha_{z}z)$ is the Hermite polynomial. 
\begin{figure*}[htbp]
%\begin{center}
\centering 

    \subfigure[]
   {\label{fig:12sff1}\includegraphics[width=0.32\linewidth]{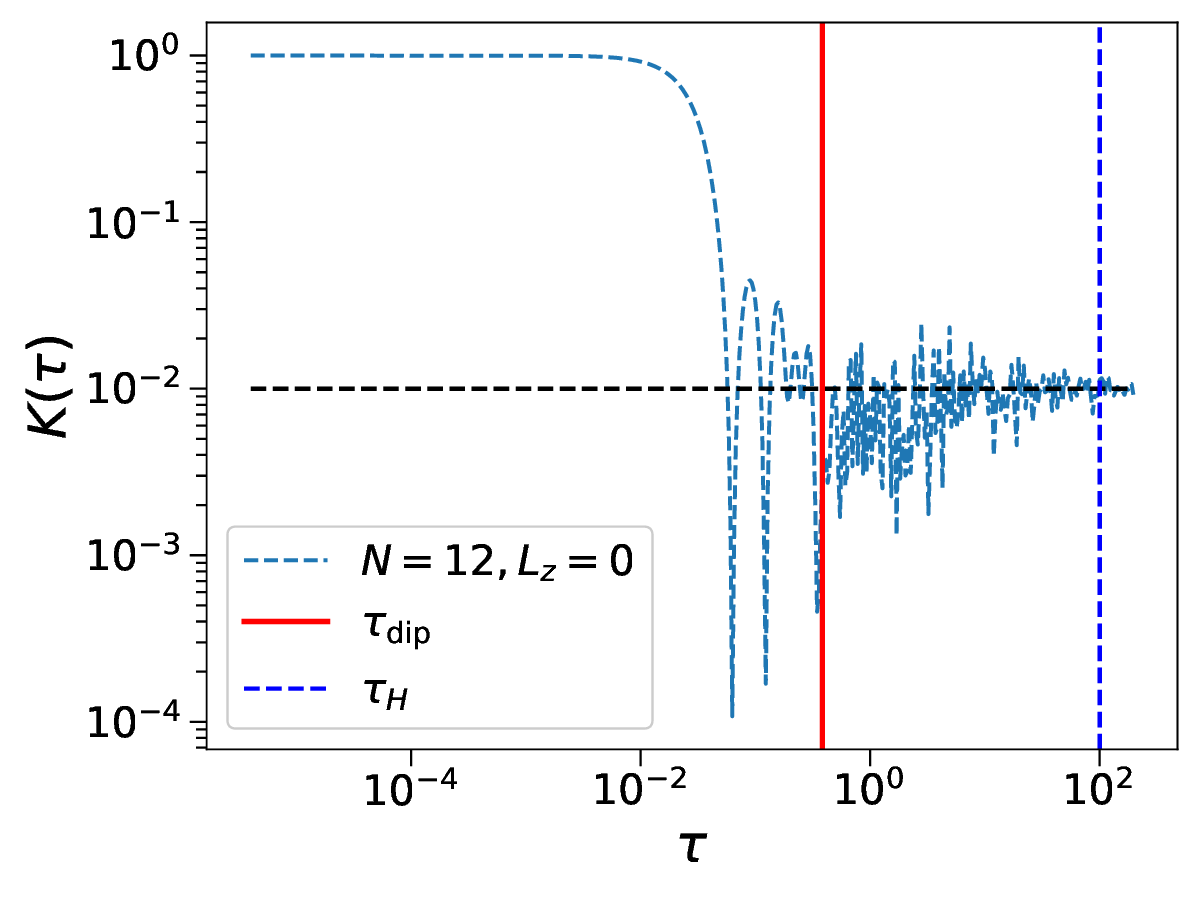}} 
    \subfigure[]
    {\label{fig:16sff1}\includegraphics[width=0.32\linewidth]{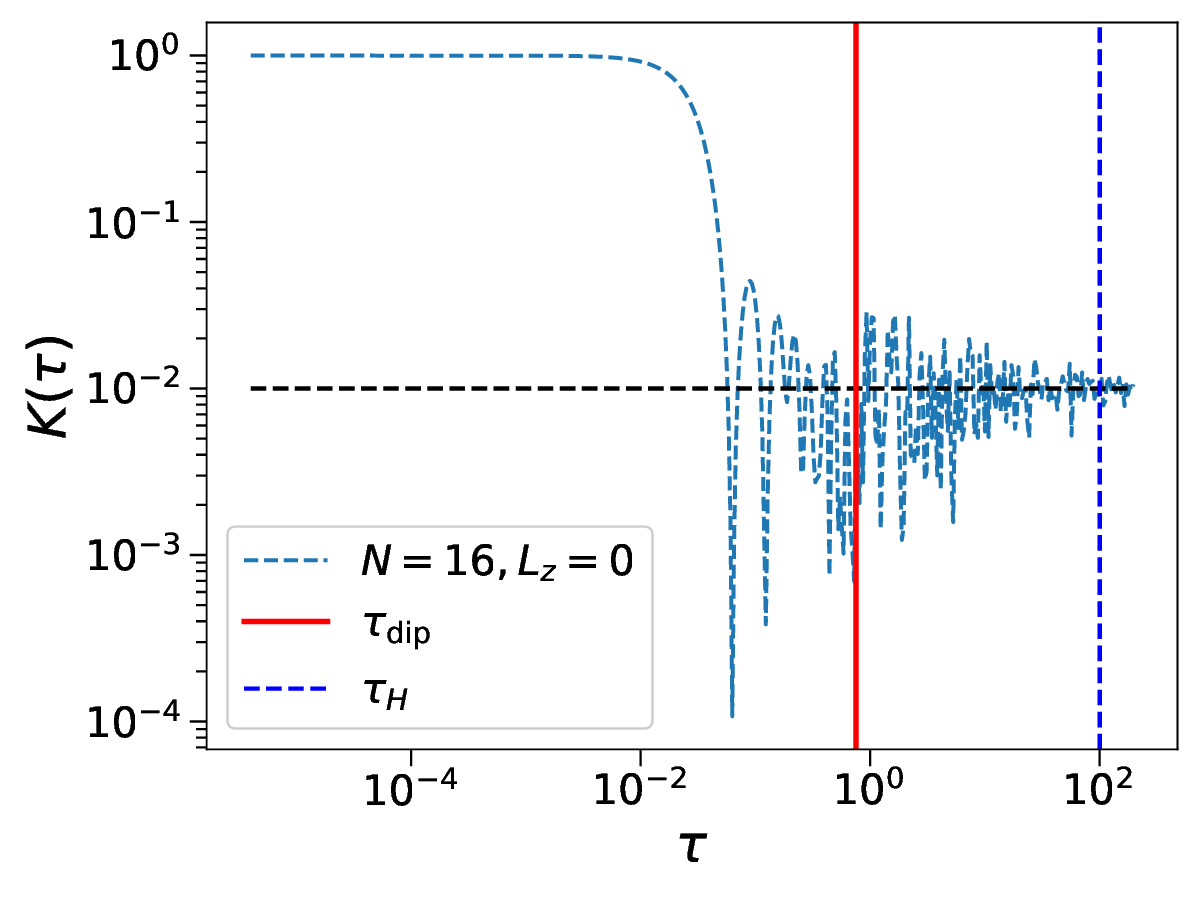}}
     \subfigure[]{\label{fig:20sff1}\includegraphics[width=0.32\linewidth]{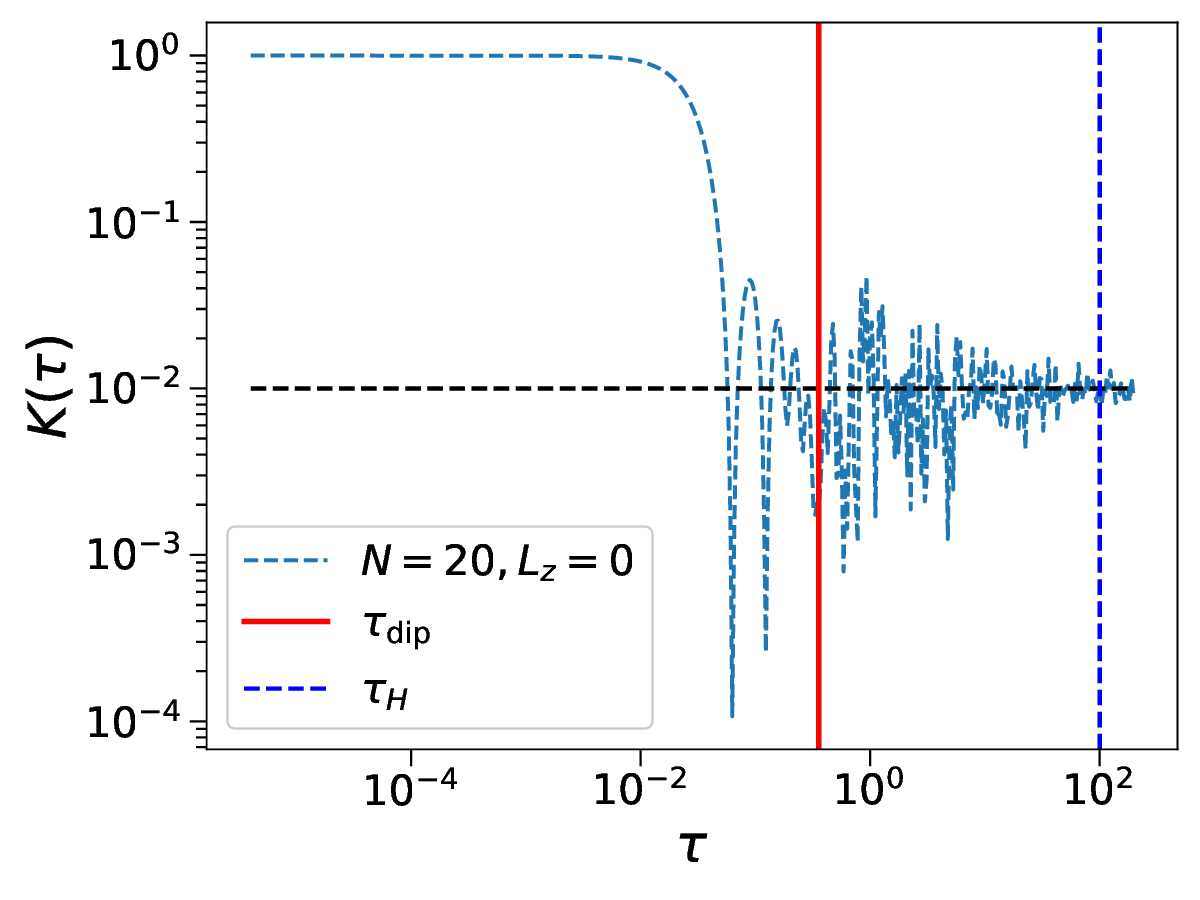}}
     \subfigure[]
   {\label{fig:12sff3}\includegraphics[width=0.32\linewidth]{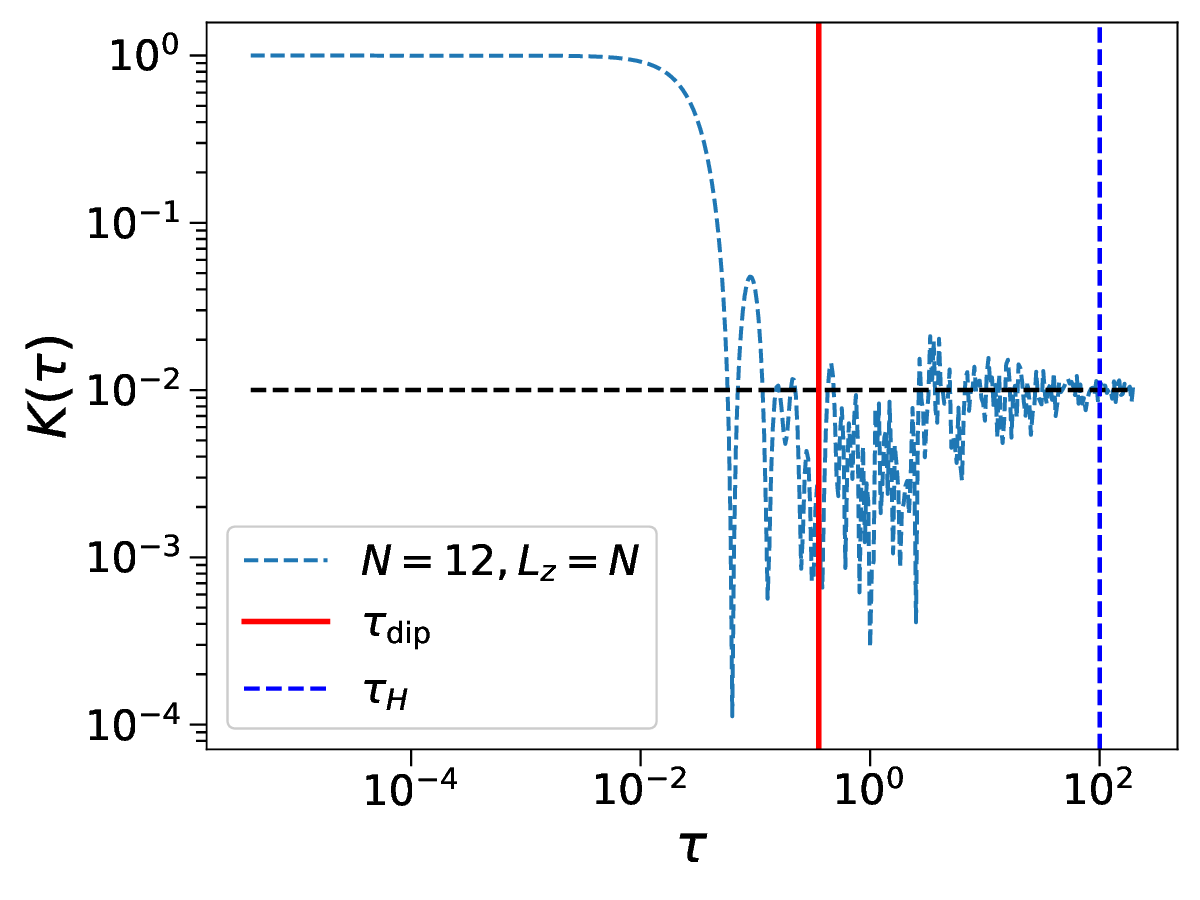}} 
    \subfigure[]
    {\label{fig:16sff3}\includegraphics[width=0.32\linewidth]{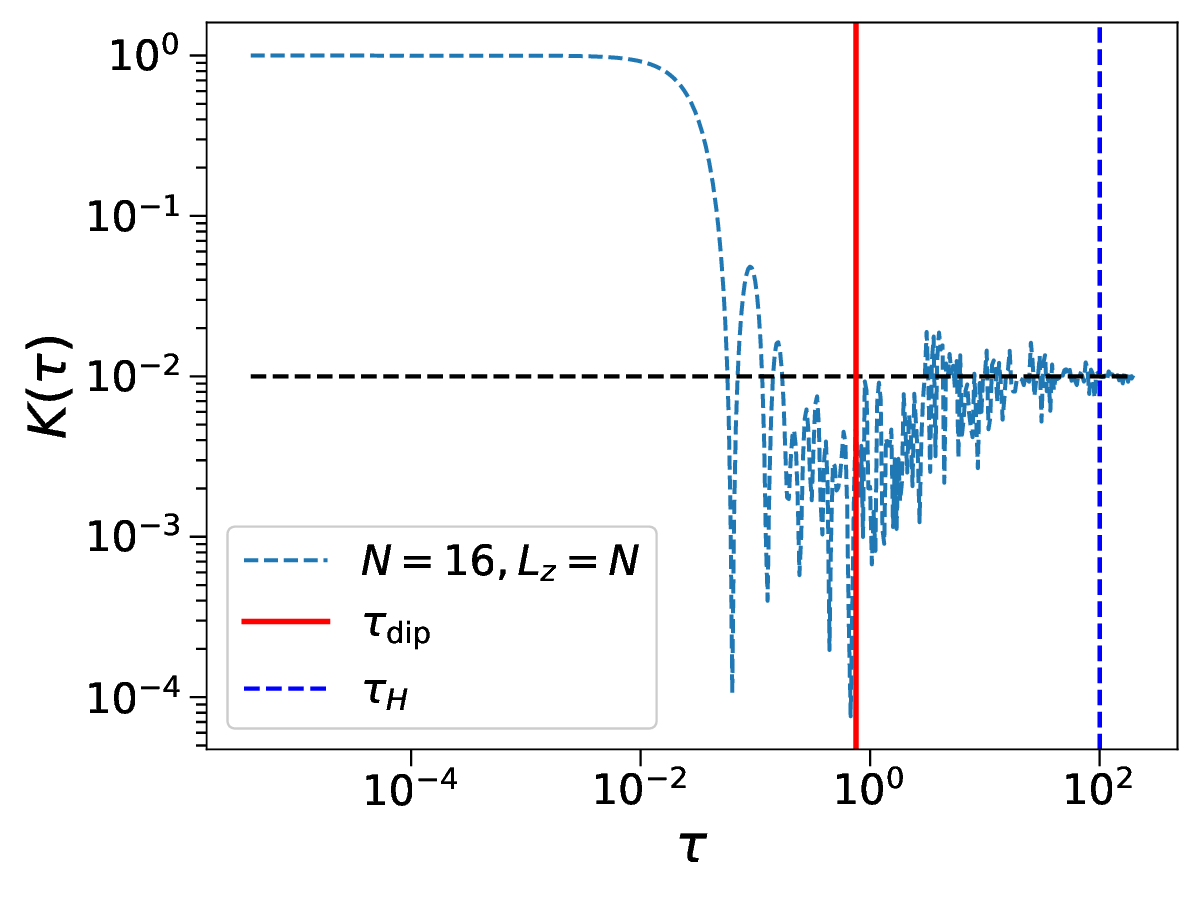}}
     \subfigure[]{\label{fig:20sff3}\includegraphics[width=0.32\linewidth]{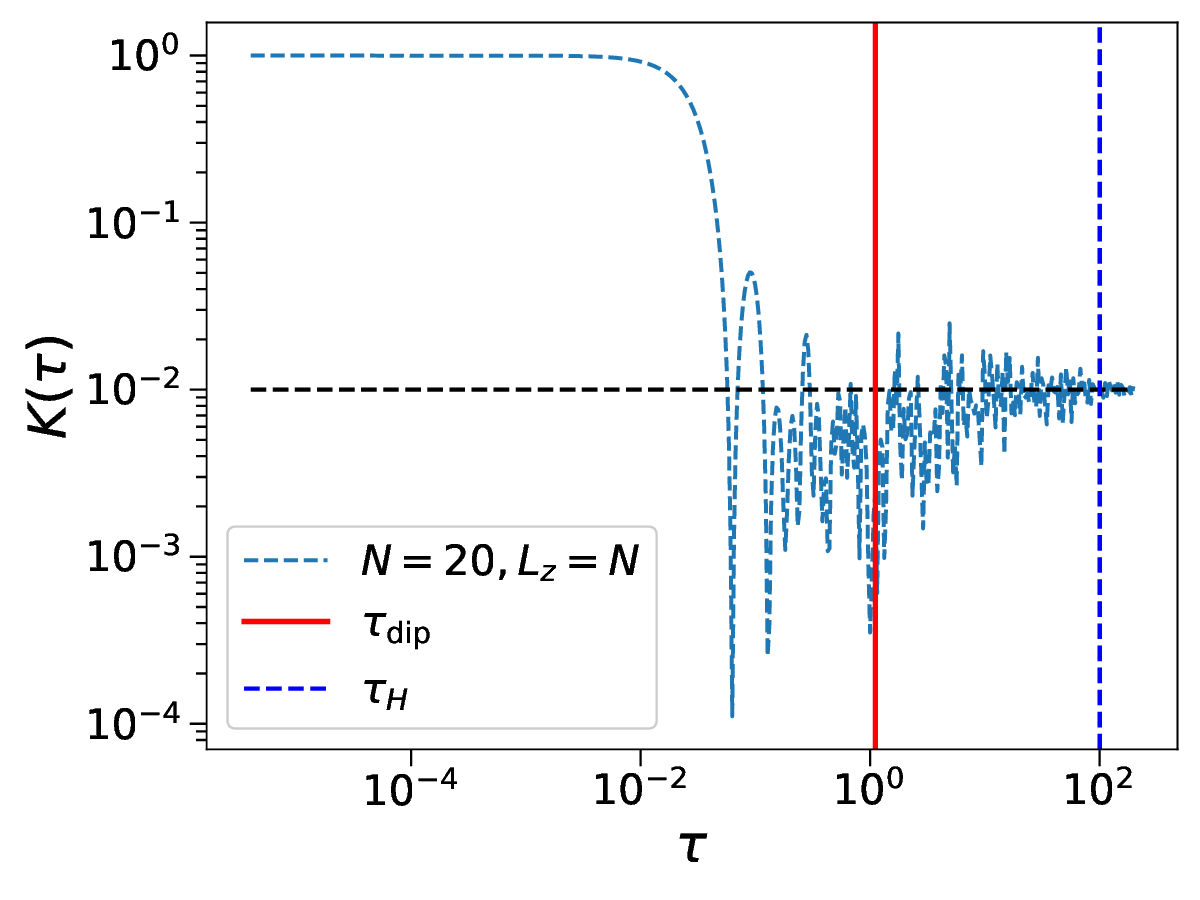}}
     
    \caption{The moderate interaction regime: Spectral form factor $K(\tau)$ \emph{vs} time, with moving average in logarithmic time window, on the log-log scale for the non-rotating case (upper panel) and the single-vortex state (lower panel) for $N=12, 16$ and $20$ bosons, with number of energy levels $\mathcal{M}=100$ utilized. The horizontal dashed line corresponds to the asymptotic limit of the SFF, $\langle K(\tau) \rangle = 1/\mathcal{M}$. The red solid and the blue dotted vertical lines mark the dip time $\tau_{dip}$ and  the Heisenberg time $\tau_{H}$, respectively. }
    \label{fig:sff1}
\end{figure*}
\begin{figure*}[htbp]
    \centering
    \includegraphics[width=0.96\textwidth]{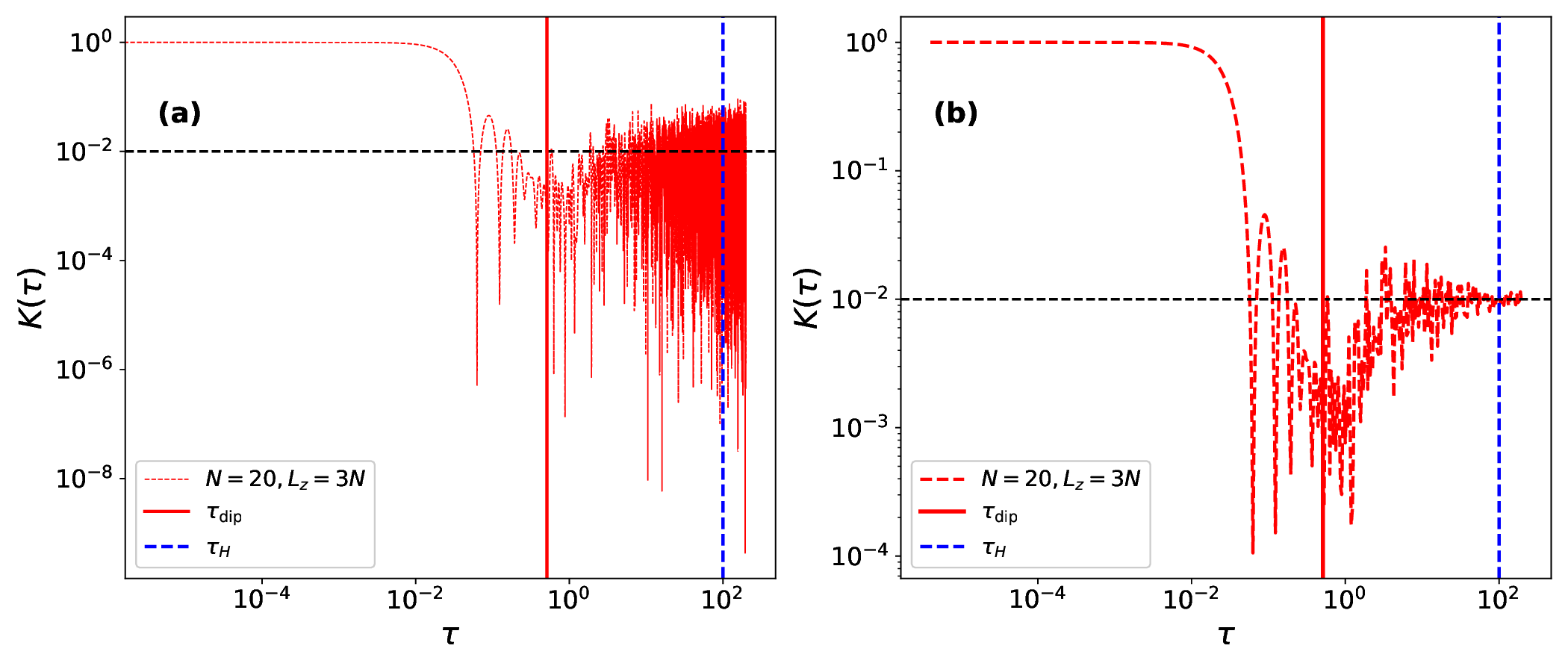}
   \caption{Spectral form factor $K(\tau)$ \emph{vs} time for $N=20$ with $L_z = 3N$. 
    (a) Raw SFF showing strong fluctuations characteristic of the non-self-averaging nature of the SFF. 
    (b) Logarithmically averaged SFF, revealing the characteristic dip-ramp-plateau structure. 
    The horizontal black dashed line marks the plateau at $\langle K(\tau) \rangle = 1/\mathcal{M}$. 
    The vertical red solid and blue dashed lines indicate the dip time $\tau_{\mathrm{dip}}$ and the Heisenberg time $\tau_H$, respectively.}
    
    \label{fig:sffraw}
\end{figure*}
\begin{figure*}[htbp]
%\begin{center}
\centering 

    \subfigure[Non-rotating]
   {\label{fig:ps1}\includegraphics[width=0.48\linewidth]{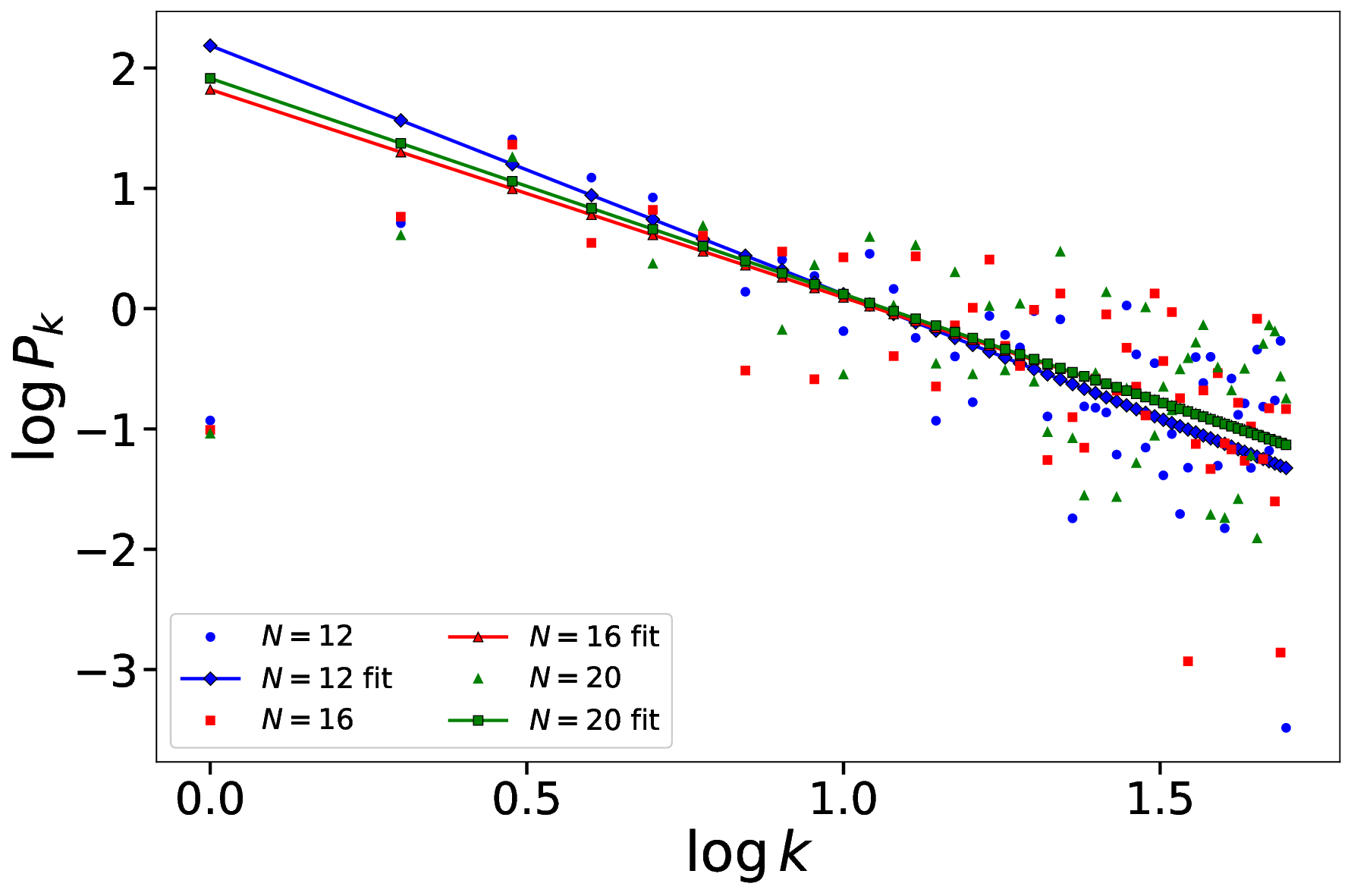}} 
    \subfigure[Single-vortex]
   {\label{fig:ps3}\includegraphics[width=0.48\linewidth]{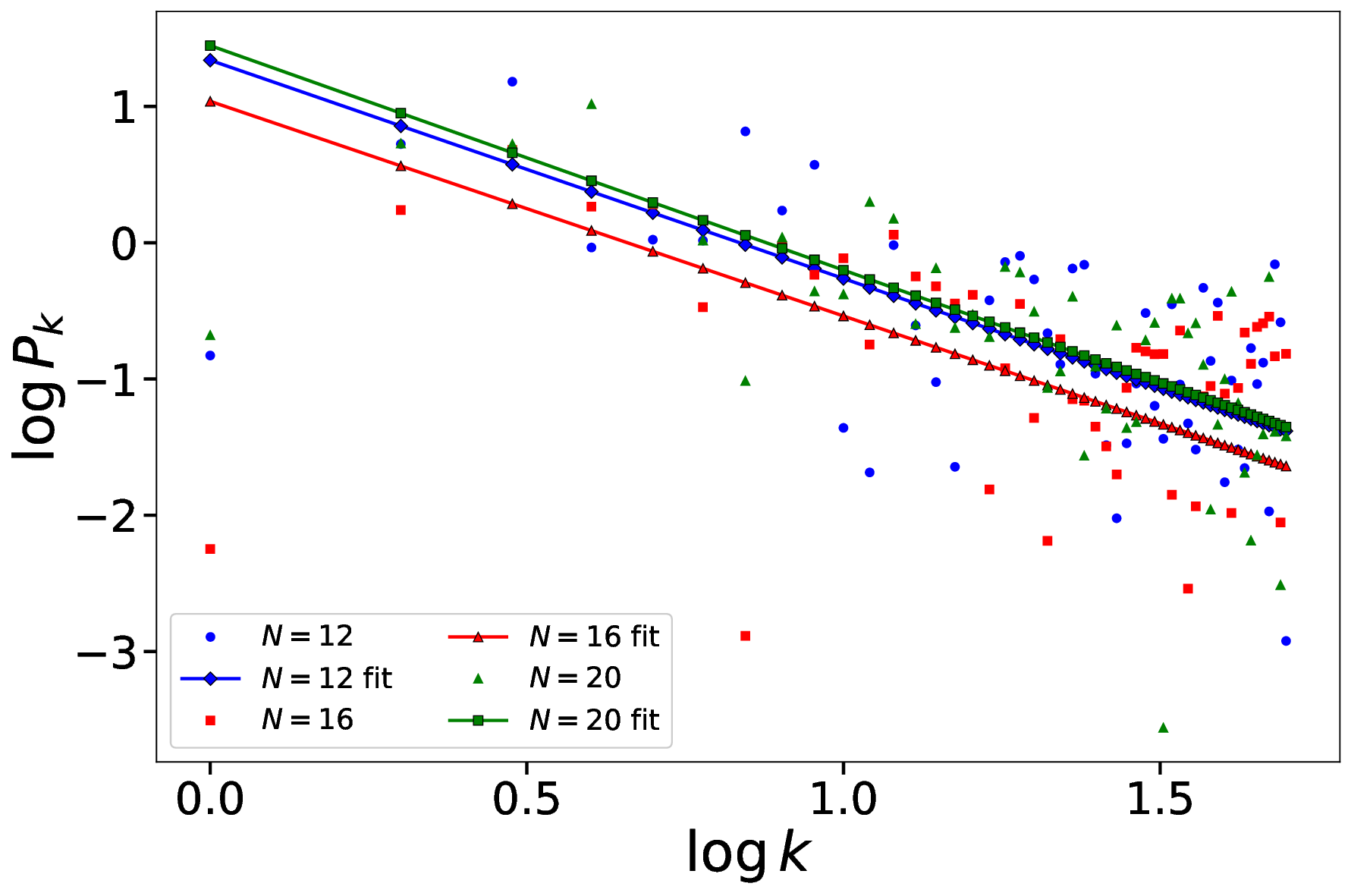}}

    \caption{The moderate interaction regime: Power spectrum $P_{k}$ vs $k$ on the decimal log-log scale in the non-rotating (left) and the single-vortex state (right) for $N=12, 16$ and $20$ bosons. The blue diamond, the red triangle and the green square lines are the straight line fits for our numerical data, denoted by dots, for $N=12, 16$ and $20$ bosons, respectively.}
    \label{fig:ps1}
\end{figure*}

\subsection{Effective Hamiltonian in two dimension}
In ultracold trapped atomic gases, it is possible to tune confining frequency and hence experimentally realize an effective two-dimensional system in the laboratory \cite{Dyke2011, Makhalov2014}. Strong axial confinement is achieved when the excitation energy $\hbar\omega_z$ along z-direction exceeds all other relevant energy scales in the system, leading to oblate-shaped traps. Thus, for $\lambda_{z} \equiv\frac{\omega_z}{\omega_{\perp}}\gg 1$, particles are constrained to move in the xy-plane with no excitation along the z-direction. The effective Hamiltonian for the system is obtained by tracing out the z-degree of freedom from the Hamiltonian in Eq. (\ref{hlab}), which with $n_{z}=0$ becomes \cite{hamid2022}

\begin{align}
\hat{H}' 
= &\sum_{i=1}^{N}\!\left[\frac{1}{2M}\!\left(\frac{\hbar}{i}\nabla_{\perp i}\right)^{2} 
+ \frac{1}{2}M\omega_{\perp}^{2}r_{\perp i}^{2}\right]
+ \frac{1}{2}\hbar \omega_{z} \nonumber\\
&+\underbrace{\frac{4\pi \hbar^{2} a_{sc}}{M} 
\, \frac{1}{\sqrt{2\pi}} 
\sqrt{\frac{\lambda_{z}}{a_{\perp}^{2}\!\left[1 + \left(\tfrac{\sigma}{a_{\perp}}\right)^{2}\! \lambda_{z}\right]}}}_{g_{2}^{\prime}} \nonumber\\
&\times \left( \frac{1}{\sqrt{2\pi}\sigma}\right)^{2}
  \sum_{i < j}^{N}
  \exp\!\left[-\frac{(r_{\perp i}-r_{\perp j})^{2}}{2\sigma^{2}}\right],
\label{hprime}
\end{align} 
where $a_{\perp} \equiv \sqrt{\frac{\hbar}{M\omega_{\perp}}}$ is the trap length and $g_{2}'$ is the effective two-body interaction strength in xy-plane having the dimension of [energy $\mathrm {\times}$ area].  In the limit of the interaction range being small compared to the trap length i.e $\tfrac{\sigma}{a_{\perp}} \ll 1$, $g_{2}'$ reduces to

\begin{align*}
    g_{2}'\approx\frac{4\pi \hbar^2 a_{sc}}{Ma_{\perp}}\sqrt{\frac{\lambda_{z}}{2\pi}}.
\end{align*}
The dimensionless effective Hamiltonian in the xy-plane in the laboratory frame defined as $\hat{H}^{\mathrm{lab}}\equiv \frac{1}{\hbar \omega_{\perp}}\hat{H}'- \frac{1}{2}\lambda_{z}$ and the angular momentum are given by
\begin{align}\label{eq4}
\hat{H}^{\mathrm{lab}}= &\sum_{i=1}^{N}\left[\frac{1}{2}\left(\frac{a_{\perp}\nabla_{\perp i}}{i}\right)^{2} 
+ \frac{1}{2}\left(\frac{r_{\perp i}}{a_{\perp}}\right)^2 \right] \nonumber \\
&+
\underbrace{g_{2}\left( \frac{a_{\perp}}{\sqrt{2\pi}\sigma}\right)^{2}\sum_{i < j}^{N}\exp\!\left[-\frac{(r_{\perp i}-r_{\perp j})^{2}}{2\sigma^{2}}\right]}_{\sum_{i < j}^{N} V_{ij}},
\end{align}

\begin{align*}
\hat{L}_{z}^{\mathrm{lab}}
= \frac{1}{i}\sum_{i=1}^{N}
\left(\mathbf{r}_{i}\times\nabla_{i}\right)_{z},
\end{align*}

\noindent where the dimensionless effective interaction strength in the xy-plane is given by
\begin{equation} \label{g2}
     g_{2} \equiv \frac{4\pi a_{sc}}{a_{\perp}} \sqrt{\frac{\lambda_{z}}{2\pi}}.
\end{equation} 
All lengths in the system are measured in the unit of the trap length, $a_{\perp}$. The first and the second term, in Eq. (\ref{eq4}), are the kinetic and the trap energies of the bosons, respectively. The third term models the two-body interaction potential by a Gaussian of width $\sigma$, which measures the range of the interaction, and $g_{2}$ is the dimensionless effective interaction strength in the xy-plane. As the system is subjected to an externally impressed rotation with angular velocity $\Omega$ about the z-axis, the Hamiltonian in the co-rotating frame is given by \cite{Ahsan2001, talib2025}
\begin{equation*}
    \hat{H}^{\mathrm{rot}} = \hat{H}^{\mathrm{lab}} - \Omega \hat{L}_{z}^{\mathrm{lab}},
\end{equation*}
where $\Omega$, measured in units of $\omega_{\perp}$,  serves as the Lagrange multiplier to constrain the system in total angular momentum state $\hat{L}_{z}^{\mathrm{lab}}$.
It can further be written as
\begin{equation}
\begin{aligned}
\hat{H}^{\mathrm{rot}}
&=
\sum_{i=1}^{N}
\underbrace{
\frac{1}{2m}
\left(
\frac{\hbar}{i}\nabla_{\perp i}
-
m\left(\Omega \hat{\mathbf e}_{z}\times\mathbf r_{\perp i}\right)
\right)^2
}_{\mathcal T_{i}}
\\[0.5em]
&\quad+
\sum_{i=1}^{N}
\underbrace{
\left[
\frac{1}{2}m\omega_{\perp}^{2}r_{\perp i}^{2}
-
\frac{1}{2}m
\left(
\Omega \hat{\mathbf e}_{z}\times\mathbf r_{\perp i}
\right)^2
\right]
}_{\mathcal U_{i}}
+\sum_{i<j}^{N}V_{ij}.
\end{aligned}
\end{equation}
As the rotational velocity approaches the transverse trapping frequency,
$\Omega \rightarrow \omega_{\perp}$, the centrifugal potential nearly
cancels the harmonic trapping potential  causing term $\mathcal U_{i}$ to vanish and leaving the free particle kinetic term in the co-rotating frame, driving the system into the Landau-level regime. The two-body interaction results in strongly correlated many-body quantum Hall-like states \cite{Baym2005}.

The crossover from integrability to chaos is encoded in the fluctuation of the many-body spectrum leading to spectral correlation. The dynamically frozen degrees of freedom do not contribute significantly to the low-energy many-body spectrum and hence it is anticipated that the two-dimensional system retains the relevant spectral correlation.

\subsection{Bose-Einstein condensation in trapped interacting bosons}The Bose-Einstein condensation (BEC) refers to the phenomenon in which a macroscopically large number of bosons occupy the same single-particle quantum state \cite{yang1962, anderson, ketterle1995}. In repulsive interacting Bose gas, it is well established \cite{Ahsan2001, wilkin2000, Cooper2001} that, in the absence of rotation ($L_{z}=0$), increasing the interaction strength causes bosons to scatter out of the single-particle state with angular-momentum $m=0$ to higher-energy states with $m\neq0$, leading to condensate depletion. As the system is subjected to an external rotation, it acquires quantized angular momenta at a series of critical angular velocities $\Omega_{c i}$ $(i = 1, 2, 3, \ldots)$. The angular momentum states corresponding to critical angular velocities $\{\Omega_{c i}\}$ are the stable states and the ones lying between two consecutive stable states are the metastable states, in the co-rotating frame \cite{Ahsan2001}. At the first critical angular velocity $\Omega_{c1}$, single-vortex state with quantized angular momentum $L_{z}=N$ is formed, corresponding to macroscopic occupation of the single-particle state with angular momentum $m=1$. Multi-vortex states corresponding to higher quantized angular momenta emerge at subsequent critical angular velocities \cite{hamid2022}. The emergence of these quantized vortex states further enhances the transfer of bosons out of the condensate to higher energy states in the co-rotating frame, amplifying the depletion of the condensate compared to the non-rotating case \cite{imran3, imran5}. As the number of bosons increases, the macroscopic occupation of the single-particle state increases due to the Bose-Einstein statistics.

\section{\label{2}Tools for quantum chaos}

The Hamiltonian in Eq. (\ref{eq4}) is iteratively diagonalized using Davidson algorithm \cite{DAVIDSON197587} to obtain variationally exact $\mathcal{M}$ lowest-lying many-body energy eigenvalues $\{ E_{i}| i=1,2,....\mathcal{M}\}$  and the corresponding eigenvectors, as described in \cite{Ahsan2001}. With the lowest $\mathcal{M}=100$ unfolded \cite{talib2025} energy levels \{$\epsilon_{i}$\}, we proceed to analyze the spectral form factor (SFF) and the power spectrum. 

\subsection{Spectral form factor}The SFF allows one to extract two important characteristic time scales of the system, namely, the dip time $\tau_{\mathrm{dip}}$ and the Heisenberg time $\tau_{H}$. The dip time $\tau_{\mathrm{dip}}$ denotes the time at which the SFF reaches its minimum value after the initial stage of oscillatory decay. This time scale signals the transition from the short-time regime, dominated by nonuniversal features and rapid fluctuations, to the onset of the linear ramp, where correlations between energy levels begin to develop and the behaviour of the SFF starts to reflect universal random-matrix predictions. In contrast, the Heisenberg time $\tau_{H}\equiv2\pi/\langle s \rangle$, with $\langle s \rangle$ denoting the mean level spacing, marks the timescale at which the linear ramp ends and SFF saturates to a plateau. In the present work, time will be measured in units of the Heisenberg time $\tau_H$. The Heisenberg time $\tau_{H}$ serves as the boundary between the linear ramp regime and the long-time plateau regime, where SFF remains constant due to finite size of the spectrum.

The spectral form factor (SFF) for a system with $\mathcal{M}$ number of energy levels is defined as \cite{mehta, stockmann, haake, Zhou_2024}

\begin{align}
    \left\langle K(\tau)\right\rangle=\left\langle \frac{|Z(i\tau)|^2}{|Z(0)|^2} \right\rangle= \frac{1}{\mathcal{M}^2} \left\langle \sum_{i,j}\exp[-i({\epsilon_{i}-\epsilon_{j}})\tau] \right\rangle,
\end{align}
where $Z(i\tau)$ is the partition function and $\tau$ is the time. The SFF is known to be a non-self-averaging quantity \cite{Prange1997} and requires appropriate averaging procedures to extract universal features. In the present study, we apply a moving average in the logarithmic time window $[\log \tau-0.018,\ \log \tau +0.018]$ \cite{patrycja,Bruno2018} to make the ramp behavior manifestly apparent. 
\begin{figure}[t]
    \centering
    \includegraphics[width=0.48\textwidth]{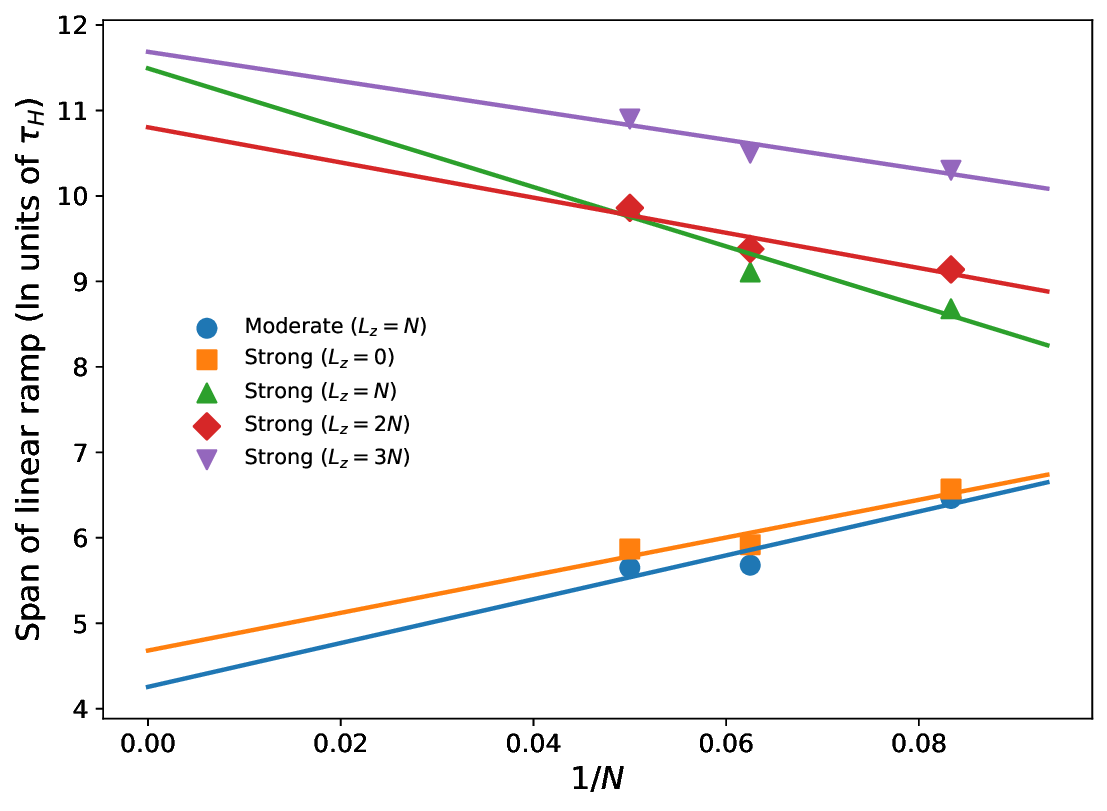}
   \caption{Finite-size scaling of the span of linear ramp with $1/N$ for the moderate interaction ($L_z=N$) and strong interaction cases ($L_z=0$, $L_z=N$, $L_z=2N$ and $L_z=3N$). The fitted lines are used to extrapolate the ramp span in the large-N limit.}
    
    \label{fig:sffscaling}
\end{figure}

\begin{table*}[htbp]
\caption{The time span of the linear ramp in the spectral form factor (SFF): In the moderate interaction regime for the non-rotating and the single-vortex state; in the strong interaction regime for the non-rotating, the single-vortex and the multi-vortex states. The span is defined as the time difference between the onset and termination of the linear ramp in SFF. The extrapolated large-$N$ limit and the corresponding slope obtained from the finite-size scaling analysis are listed in the last two rows.}
\label{sfframptable}
\centering
\begin{ruledtabular}
\begin{tabular}{c c c c c c c}
$N$ &
\multicolumn{2}{c}{Moderate} &
\multicolumn{4}{c}{Strong} \\
%\multicolumn{2}{c}{Multi-vortex (strong)} \\
\cline{2-3} \cline{4-7} %\cline{6-7}
& Non-rotating & Single-vortex ($L_z=N$) & Non-rotating & Single-vortex & $L_z=2N$ & $L_z=3N$ \\
\hline
12 & no ramp & 6.46& 6.57 & 8.68&  9.14 & 10.3 \\
16 & no ramp & 5.68& 5.92 & 9.11&  9.38 & 10.5 \\
20 & no ramp & 5.65& 5.87 & 9.89&  9.86 & 10.9 \\
\hline
Large-$N$ limit
& &4.25 & 4.68& 11.5& 10.8& 11.7 \\
\hline
Slope
& &25.6 & 22.0& -34.7& -20.6& -17.1 \\
\end{tabular}
\end{ruledtabular}
\end{table*}

To illustrate the need for moving average, we show in Fig.~\ref{fig:sffraw} both the raw and logarithmically averaged SFF. The raw SFF [Fig.~\ref{fig:sffraw}(a)] exhibits strong fluctuations that obscure the underlying universal structure. These fluctuations are a manifestation of the non-self-averaging nature of the SFF. After applying moving average in a logarithmic window, the characteristic dip-ramp-plateau structure of quantum chaotic system becomes clearly visible [Fig.~\ref{fig:sffraw}(b)]. 

To examine the finite-size effects on the spectral form factor, we perform finite-size scaling analysis for the time span of the linear ramp defined as
\begin{equation*}
L_{\rm ramp}=\tau_{\rm end}-\tau_{\rm start},
\end{equation*}
where $\tau_{\rm start}$ and $\tau_{\rm end}$ denote the onset and termination time of the linear ramp, respectively. 
The resulting values are fitted to the finite-size scaling form
\begin{equation}
L_{\rm ramp}(N)=L_{\infty}+\frac{x}{N},
\end{equation}
where $L_{\infty}$ denotes the extrapolated value of the span of the linear ramp in the large-$N$ limit, and $x$ represents the slope of the linear fit, as shown in Fig.~\ref{fig:sffscaling} and summarized in Table~\ref{sfframptable}.

For the single-vortex state ($L_z=N$) in the moderate interaction regime and the non-rotating state ($L_z=0$) in the strong interaction regime, the finite-size scaling yields a positive slope, as shown in Fig.~\ref{fig:sffscaling} and Table~\ref{sfframptable}, indicating that the span of the linear ramp decreases with increasing system size. This behavior suggests that these states evolve towards integrability due to increasing degree of Bose-Einstein condensation with increase in number of bosons. 
In contrast, the single-vortex ($L_z=N$), $L_z=2N$, and $L_z=3N$ states in the strong interaction regime exhibit negative slopes, as shown in Fig.~\ref{fig:sffscaling} and Table~\ref{sfframptable}, implying that the span of the linear ramp increases with increase in number of bosons. The trend indicates that the signatures of quantum chaos persist and become more pronounced as the number of bosons increases due to depletion of Bose-Einstein condensate caused by strong interaction and further amplified by rotation leading to nucleation of multi-vortices.

\subsection{Power spectrum}The power spectrum effectively quantifies deviation of the unfolded excitation energies from their mean value. In a quantum system, fluctuations of the energy-level spacings $s_i = \epsilon_{i+1} - \epsilon_i$ about their mean value $\langle s \rangle$ defined as
\begin{align}
 \begin{split}
     \delta_n &= \sum_{i=1}^n (s_i - \langle{s}\rangle),
 \end{split}
\end{align}
can be treated as discrete time series \cite{Relano2002, Faleiro2004, Relano2008, das2025}.
The Fourier transform of this series

\begin{equation}
  \hat{\delta}_k = \frac{1}{\sqrt{L}}\sum_{n}\delta_n e^{-\frac{2\pi i kn}{L}},
\end{equation}
and the corresponding power spectrum given by
\begin{align}
	\label{eq_P_Noise_def}
	P_k = |\hat{\delta}_k|^2,
\end{align}
provides a means of distinguishing transition from integrable to chaotic behavior \cite{Riserprl2017}. For chaotic systems, ${P_k}\propto \frac{1}{k}$, whereas integrable systems follow ${P_k}\propto \frac{1}{k^2}$. This behavior holds regardless of whether time-reversal invariance is preserved, meaning it is independent of the universality class.

\section{\label{3}Results and discussion}
We consider a system of $N = 12, 16,$ and $20$ bosonic atoms of $^{87}\mathrm{Rb}$ with trap anisotropy parameter $\lambda_{z} = \frac{\omega_{z}}{\omega_{\perp}} = 4$ and axial frequency of the trap $\omega_{z} = 2\pi \times 220~\text{Hz}$ \cite{Dalfovo1996,Baym1996}. These choices of parameters and mass $M$ of the $^{87}\mathrm{Rb}$ atom yield radial trap length of $a_{\perp} = \sqrt{\frac{\hbar}{M \omega_{\perp}}} = 1.446~\mu\text{m}$. The interaction range of the Gaussian potential is fixed at $\sigma = 0.1\,a_{\perp}$. The effective interaction strength in the mean-field approximation for contact potential is characterized by the parameter $\frac{N a_{sc}}{a_{\perp}}$ \cite{dalfovo1999}. In the present work, we restrict our study to the repulsive regime with $a_{sc} > 0$. We employ exact diagonalization method in which the exponential increase of the many-body Hilbert space with increasing number of bosons restricts the number of bosons $N$ to a few tens of bosons. To achieve values of the parameter $Na_{sc}/a_{\perp}$ relevant to experimental situations, we parametrically vary the $s$-wave scattering length in our theoretical analysis. With $a_{0} = 0.05292~\text{nm}$ as the Bohr radius, we take $a_{sc} = 1000\,a_{0}$ and $a_{sc} = 10000\,a_{0}$ to represent the moderately and the strongly interacting regime, respectively. With these parameters, the dimensionless interaction strength for the quasi-2D system, as given in Eq. (\ref{g2}), takes values $g_{2} = 0.3669$ in the moderate interaction regime and $g_{2} = 3.669$ in the strong interaction regime.

We consider two distinct interaction regimes: the moderate interaction regime where the interaction energy remains small compared to the trap energy, and the strong interaction regime where the interaction energy becomes comparable to the trap energy. The dimensionless effective two-body interaction strength parameter in Eq. (\ref{g2}) takes values $g_{2}=0.3669$ and $g_{2}=3.669$ for moderate and strong interaction regimes, respectively, as derived in the appendix. We numerically investigate the spectral form factor (SFF) and the power spectrum for varying number of bosons, considering moderate interaction regime for non-rotating, rotating single-vortex states and strong interaction regime for non-rotating, rotating single-vortex and multi-vortex states with $L_{z}=N, 2N, 3N$.

\begin{figure*}[!t]
%\begin{center}
\centering 

     \subfigure[]{\label{fig:12sff2}\includegraphics[width=0.32\linewidth]{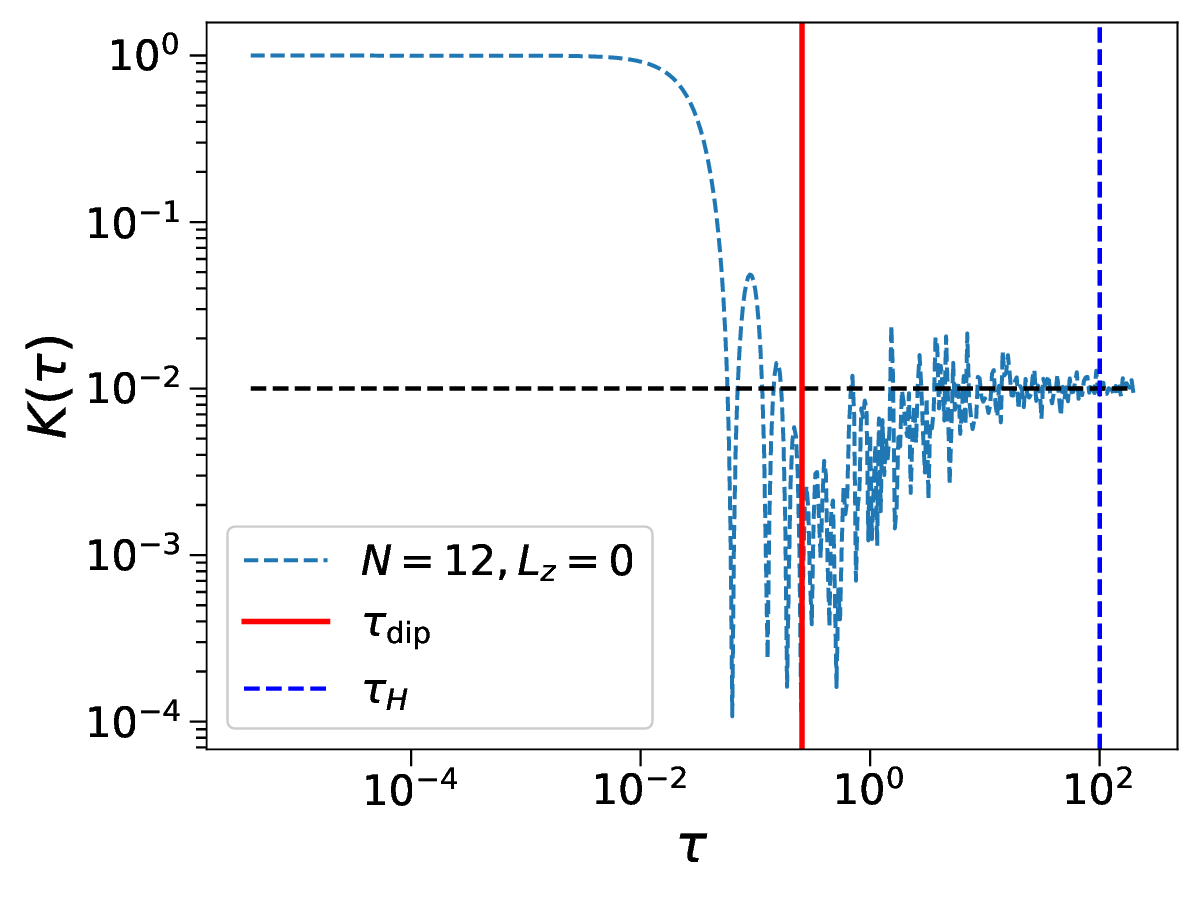}} 
     \subfigure[]{\label{fig:16sff2}\includegraphics[width=0.32\linewidth]{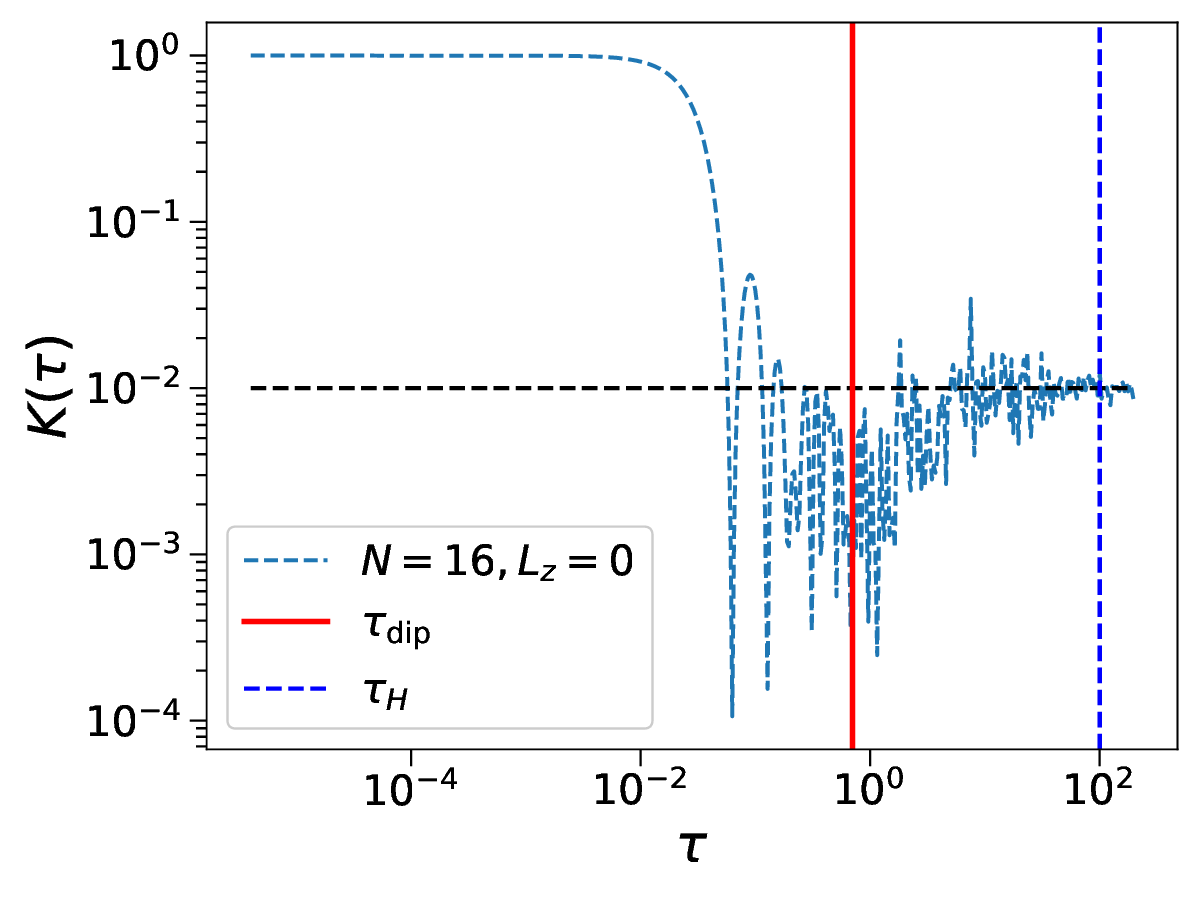}}
     \subfigure[]{\label{fig:20sff2}\includegraphics[width=0.32\linewidth]{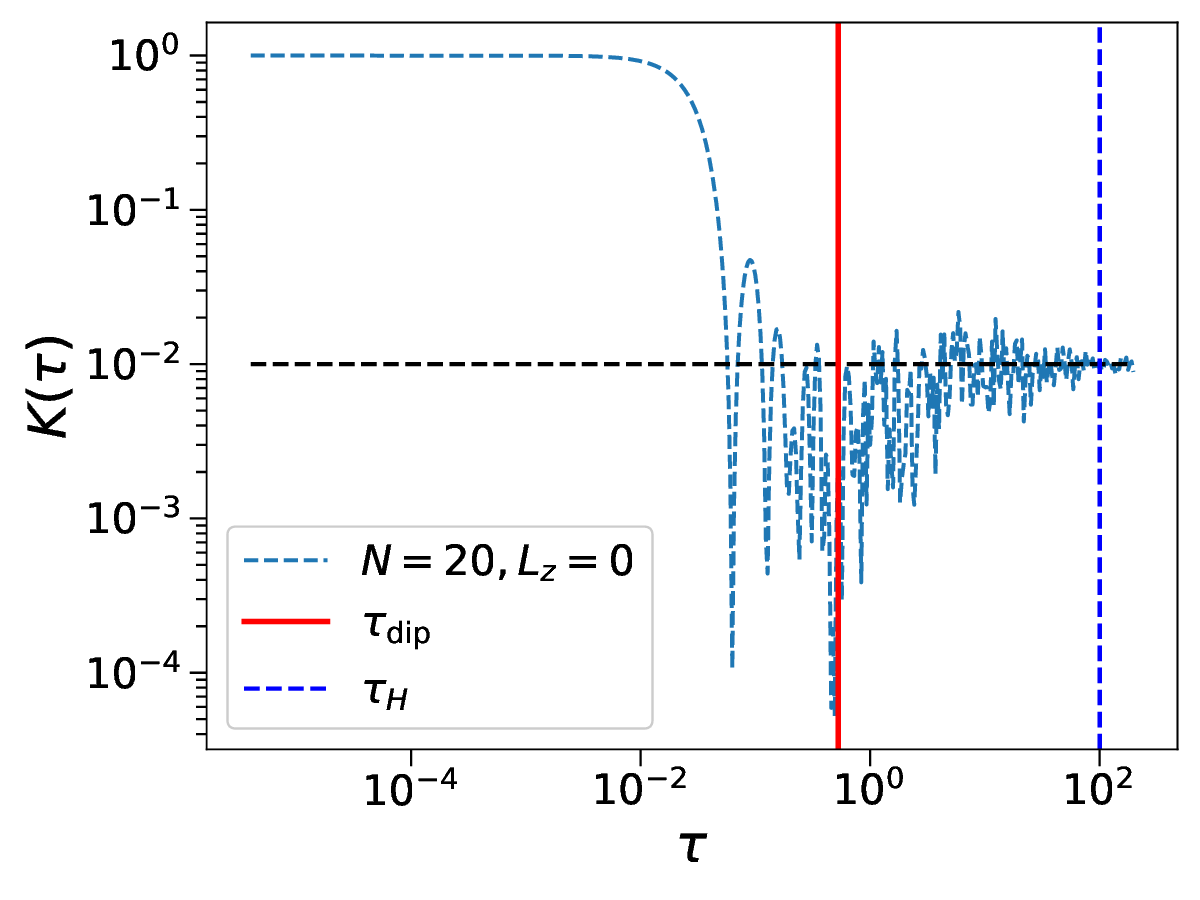}}
      \subfigure[]{\label{fig:12sff4}\includegraphics[width=0.32\linewidth]{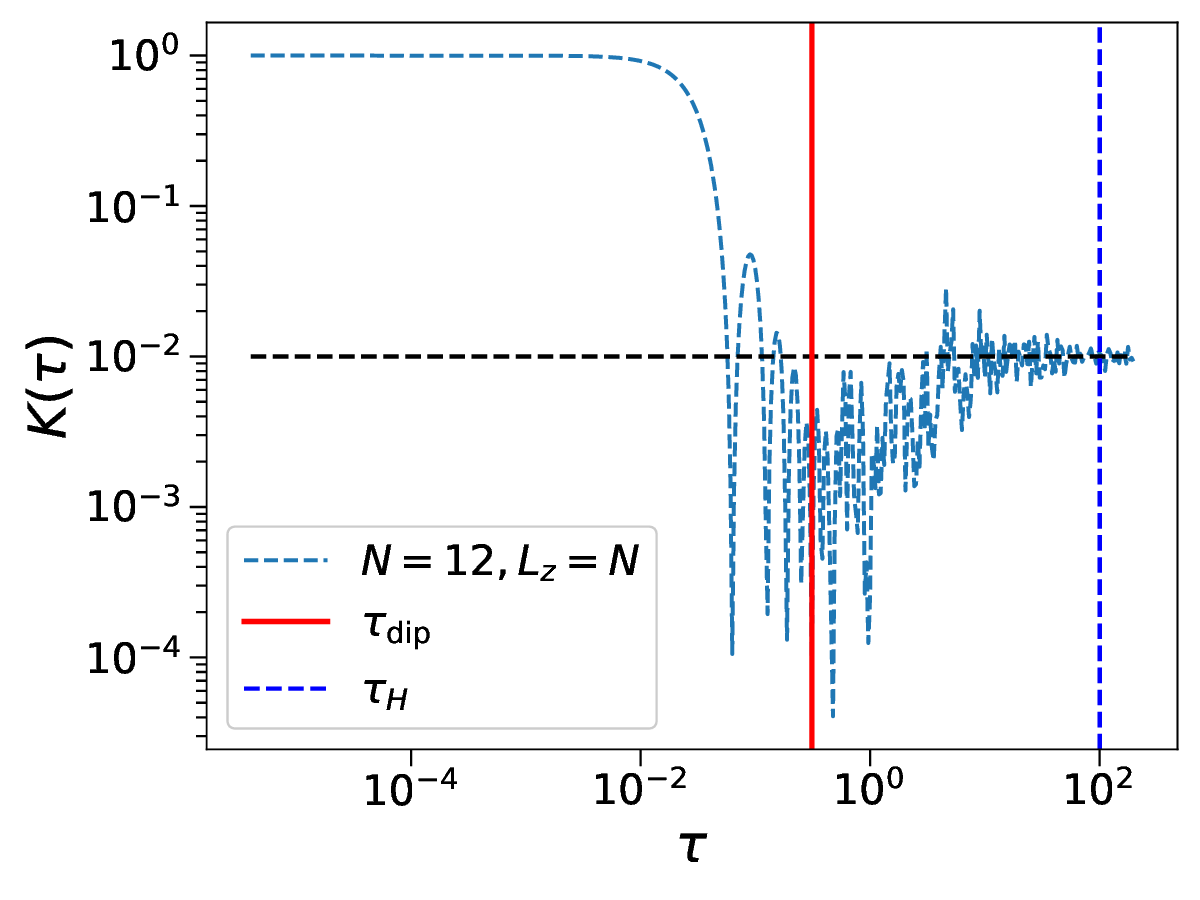}}
      \subfigure[]{\label{fig:16sff4}\includegraphics[width=0.32\linewidth]{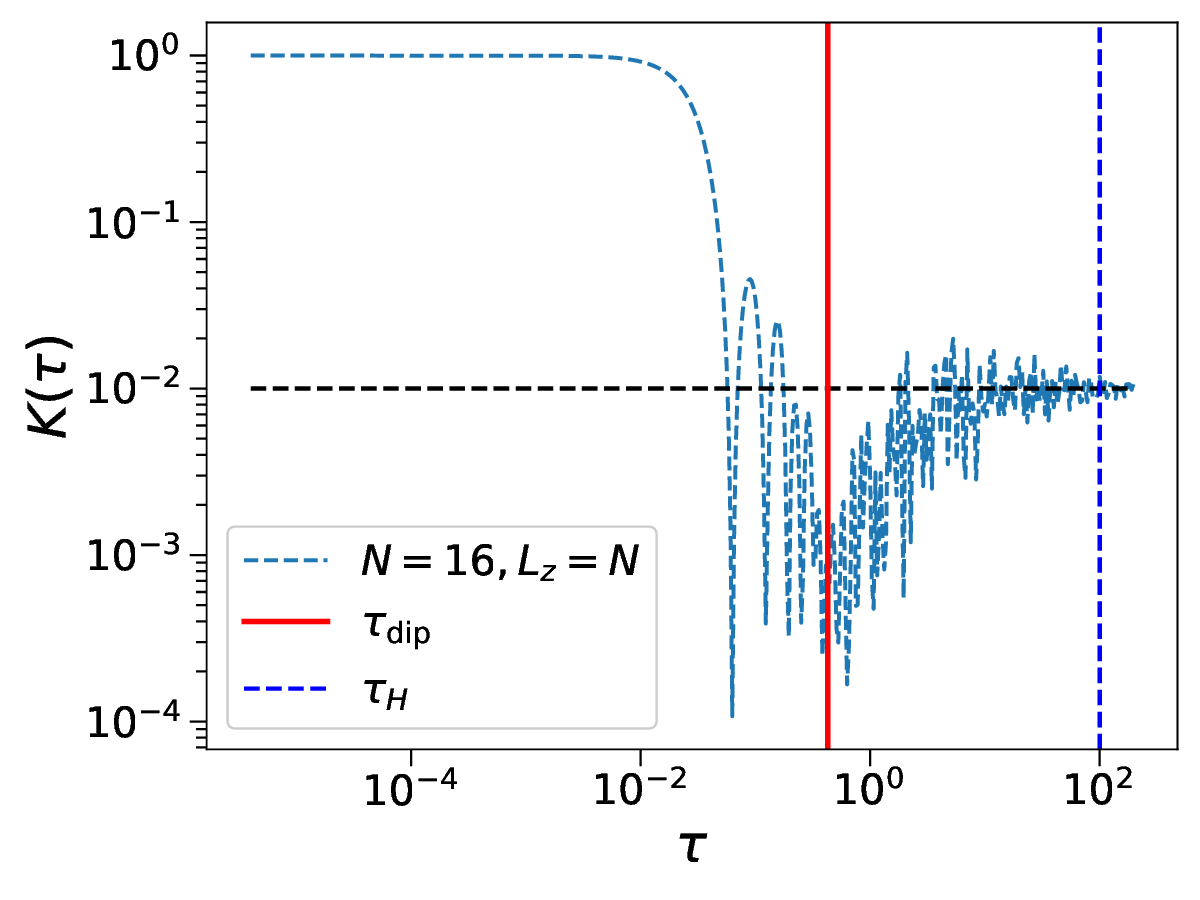}}
      \subfigure[]{\label{fig:20sff4}\includegraphics[width=0.32\linewidth]{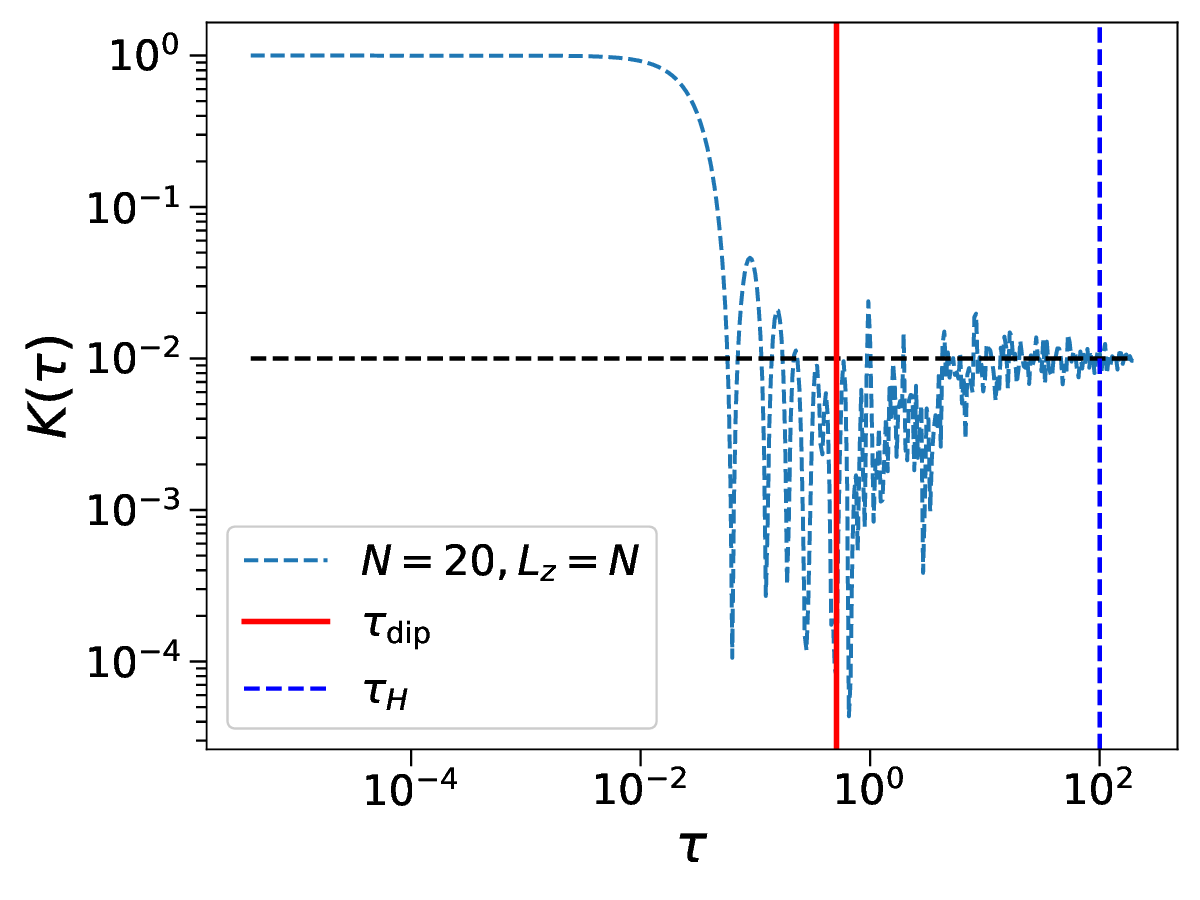}}
       \subfigure[]{\label{fig:12sff5}\includegraphics[width=0.32\linewidth]{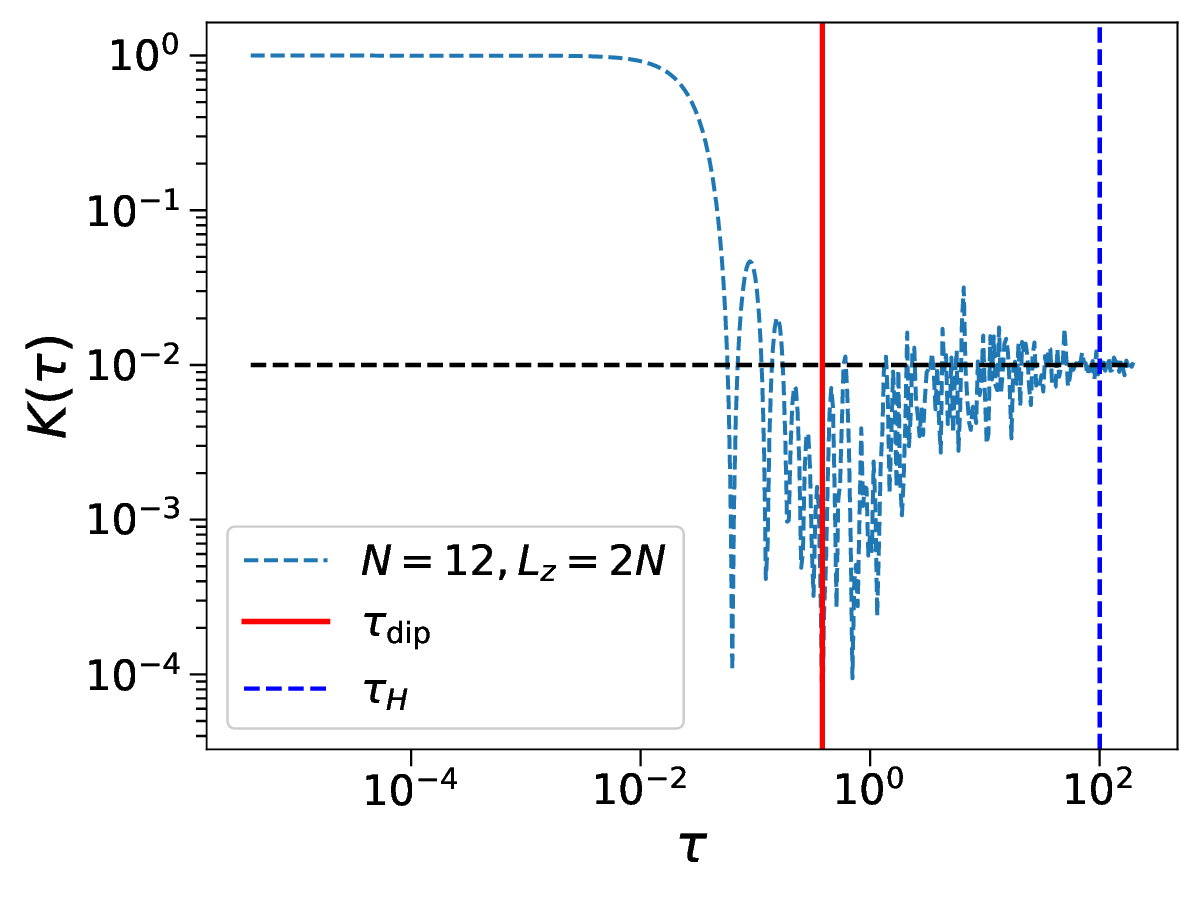}} 
       \subfigure[]{\label{fig:16sff5}\includegraphics[width=0.32\linewidth]{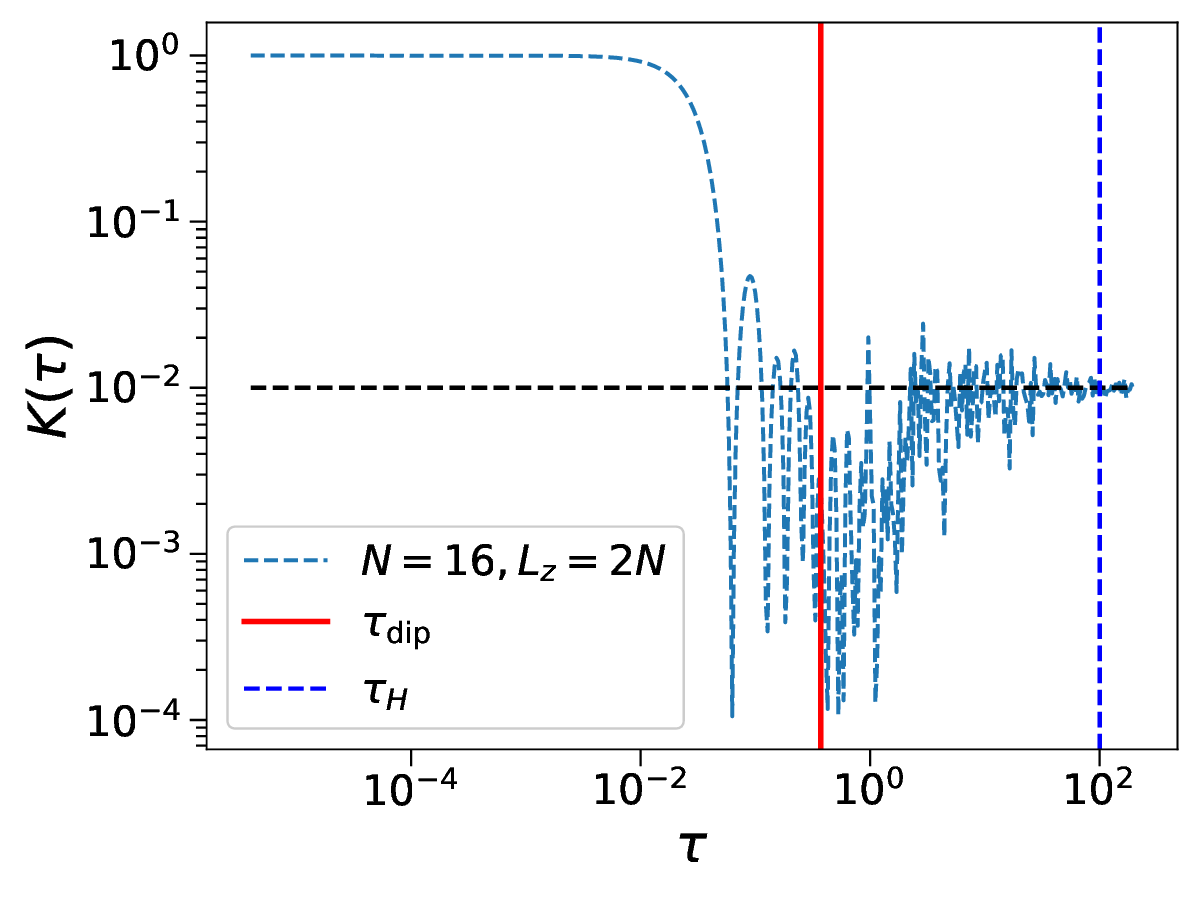}}
       \subfigure[]{\label{fig:20sff5}\includegraphics[width=0.32\linewidth]{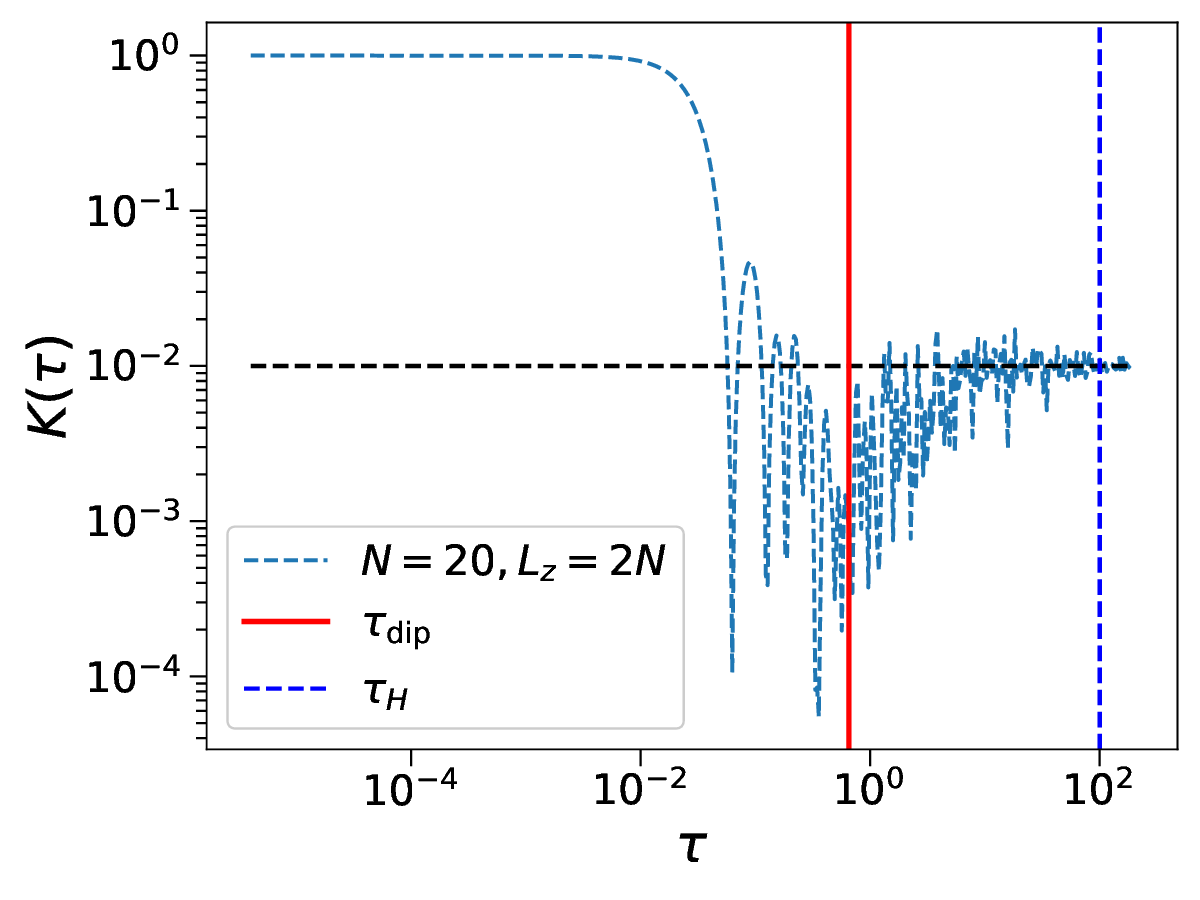}}
       \subfigure[]{\label{fig:12sff6}\includegraphics[width=0.32\linewidth]{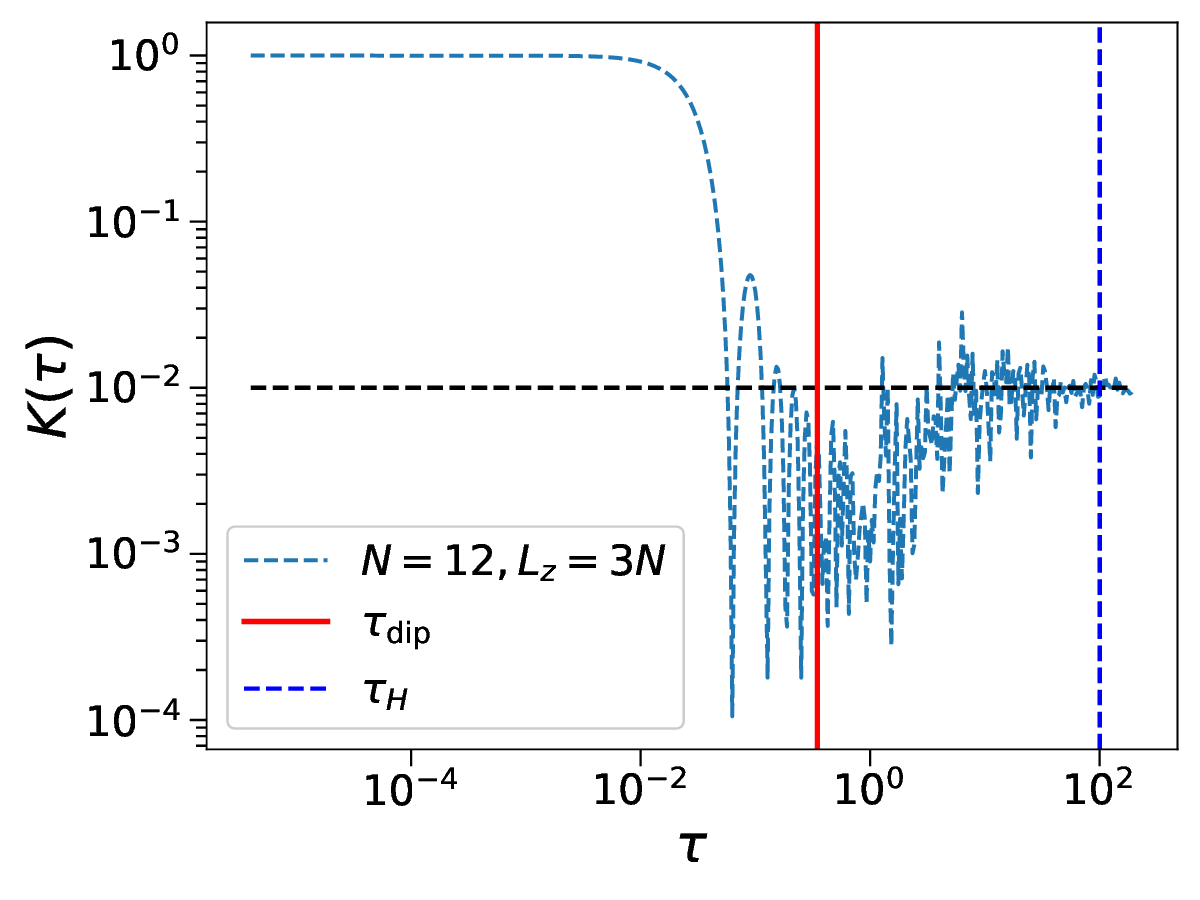}}
       \subfigure[]{\label{fig:16sff6}\includegraphics[width=0.32\linewidth]{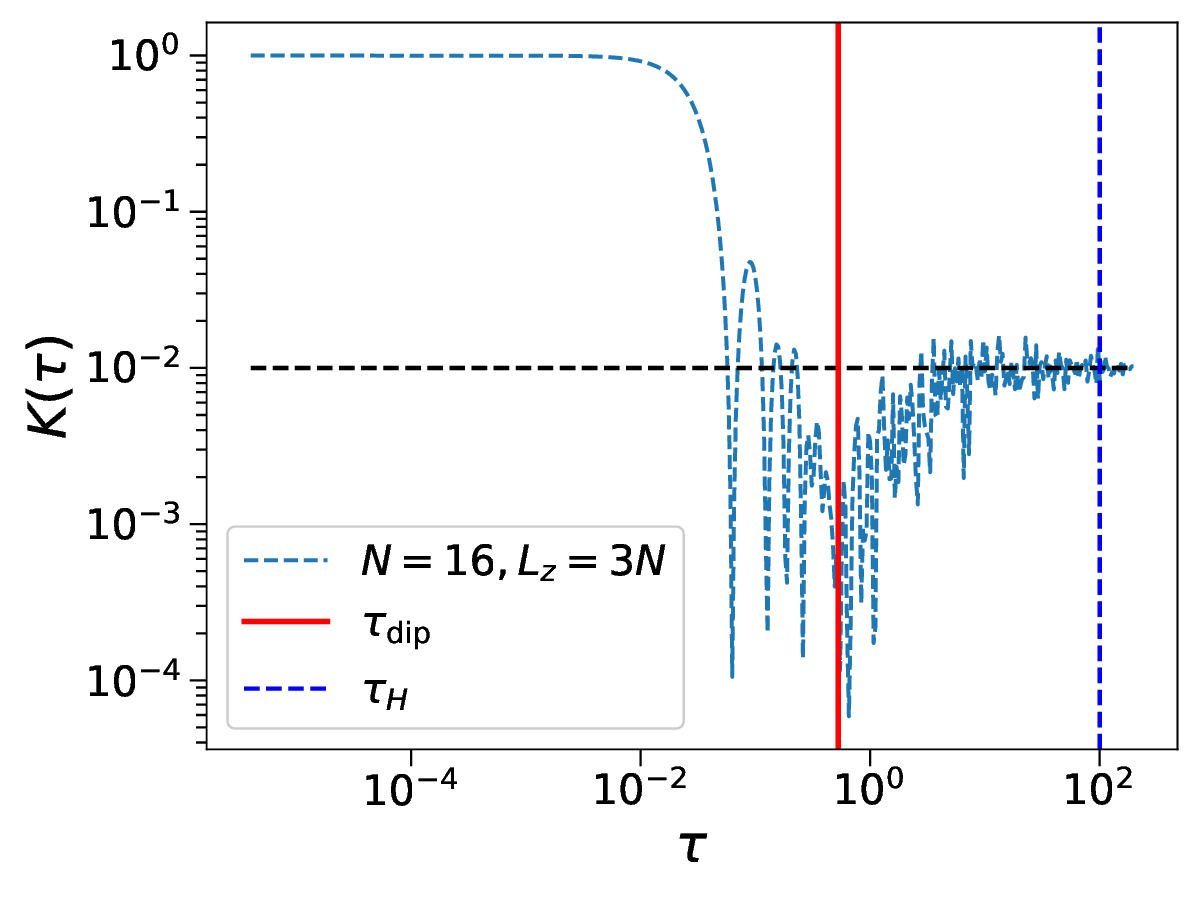}}  
       \subfigure[]{\label{fig:20sff6}\includegraphics[width=0.32\linewidth]{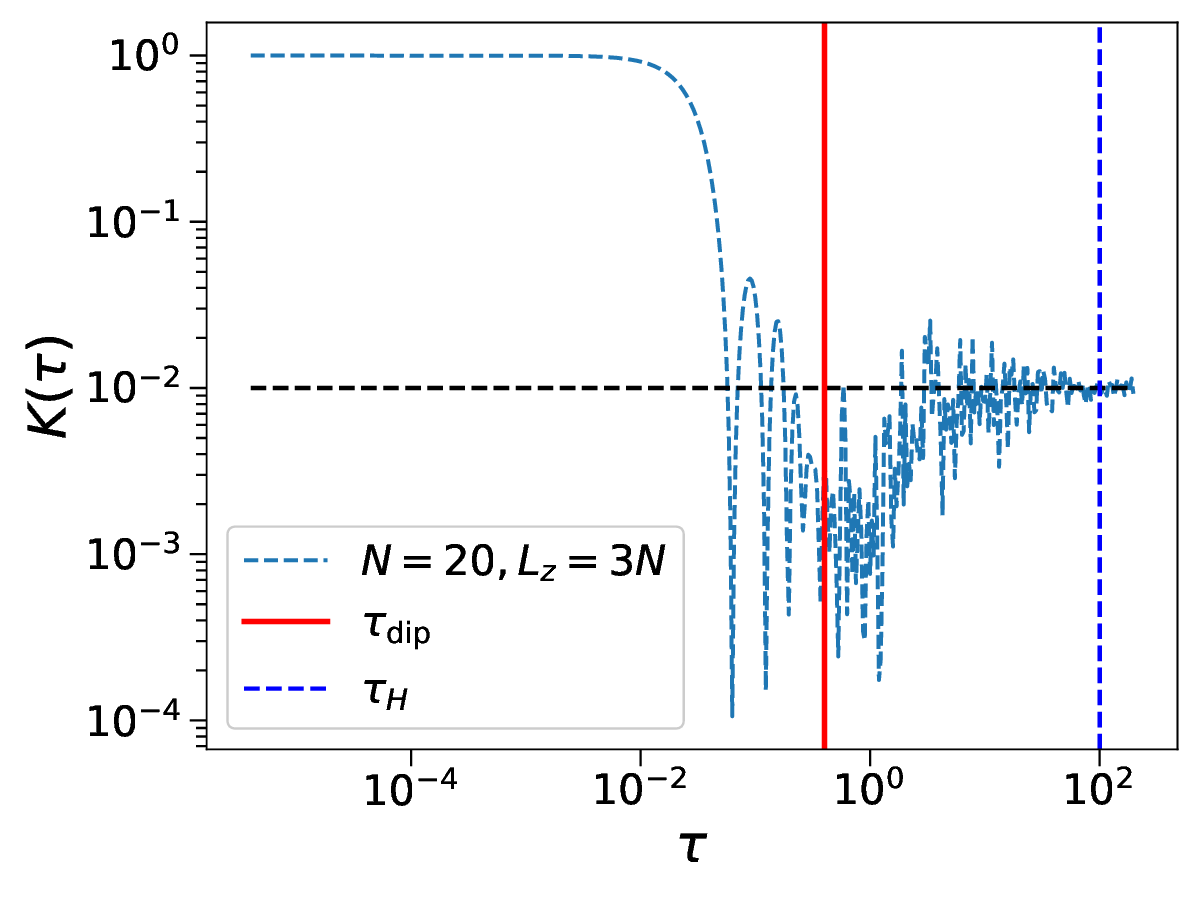}}
     
    \caption{The strong interaction regime: Spectral form factor $K(\tau)$ \emph{vs} time, with moving average in logarithmic time window, on the log-log scale in the non-rotating case (first row), the single-vortex state (second row), the multi-vortex states $L_{z}=2N$ (third row) and $L_{z}=3N$ (fourth row) for $N=12, 16$ and $20$ bosons, with number of energy levels $\mathcal{M}=100$ utilized. The horizontal dashed line corresponds to the asymptotic limit of the SFF, $\langle K(\tau) \rangle = 1/\mathcal{M}$. The red solid and the blue dotted vertical lines mark the dip time $\tau_{dip}$ and  the Heisenberg time $\tau_{H}$, respectively.}
    \label{fig:sff2}
\end{figure*}

\begin{figure*}[!t]
\centering

   \subfigure[Non-rotating]
    {\label{fig:ps2}\includegraphics[width=0.48\linewidth]{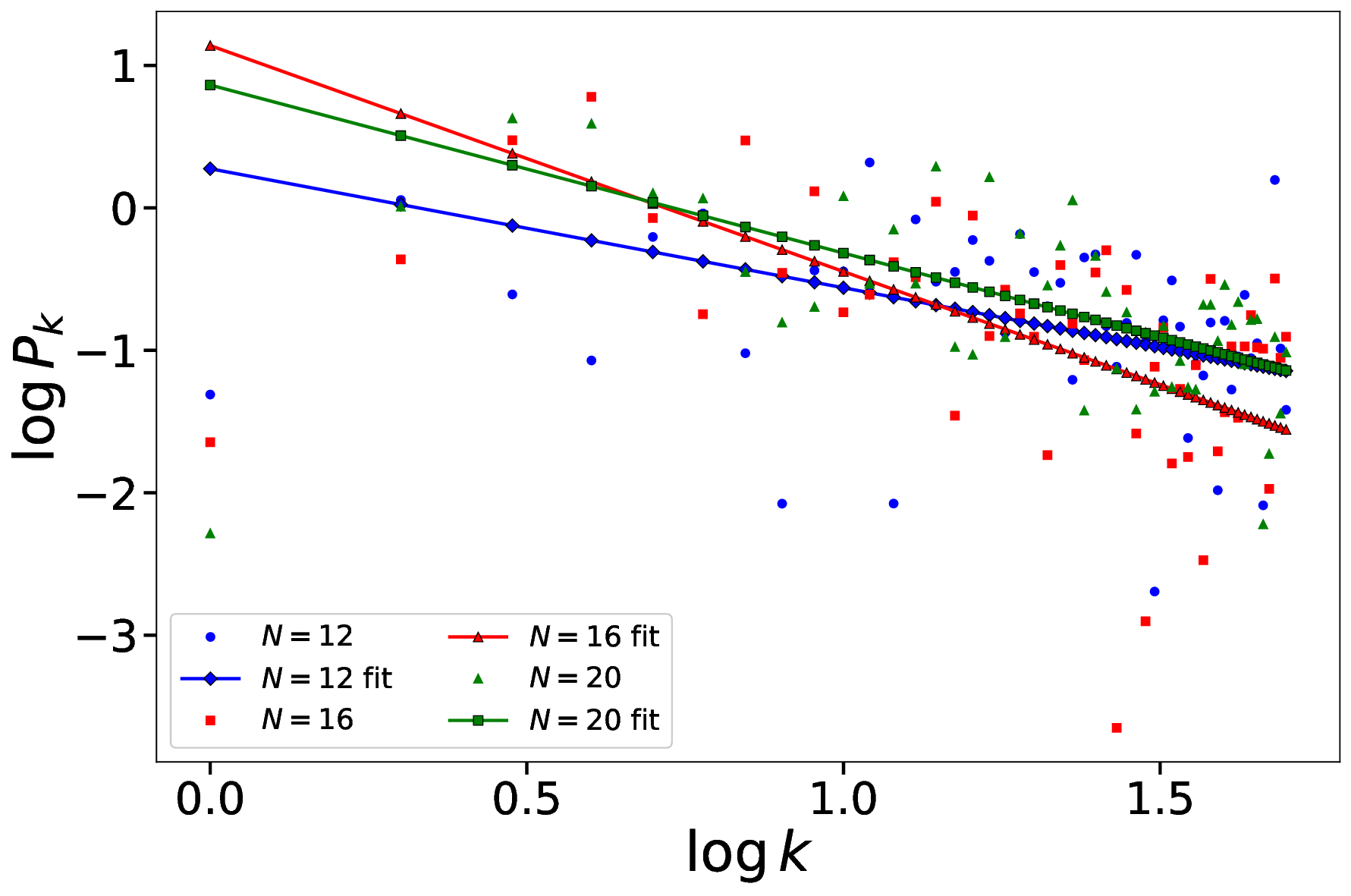}}
   \subfigure[Single-vortex]
    {\label{fig:ps4}\includegraphics[width=0.48\linewidth]{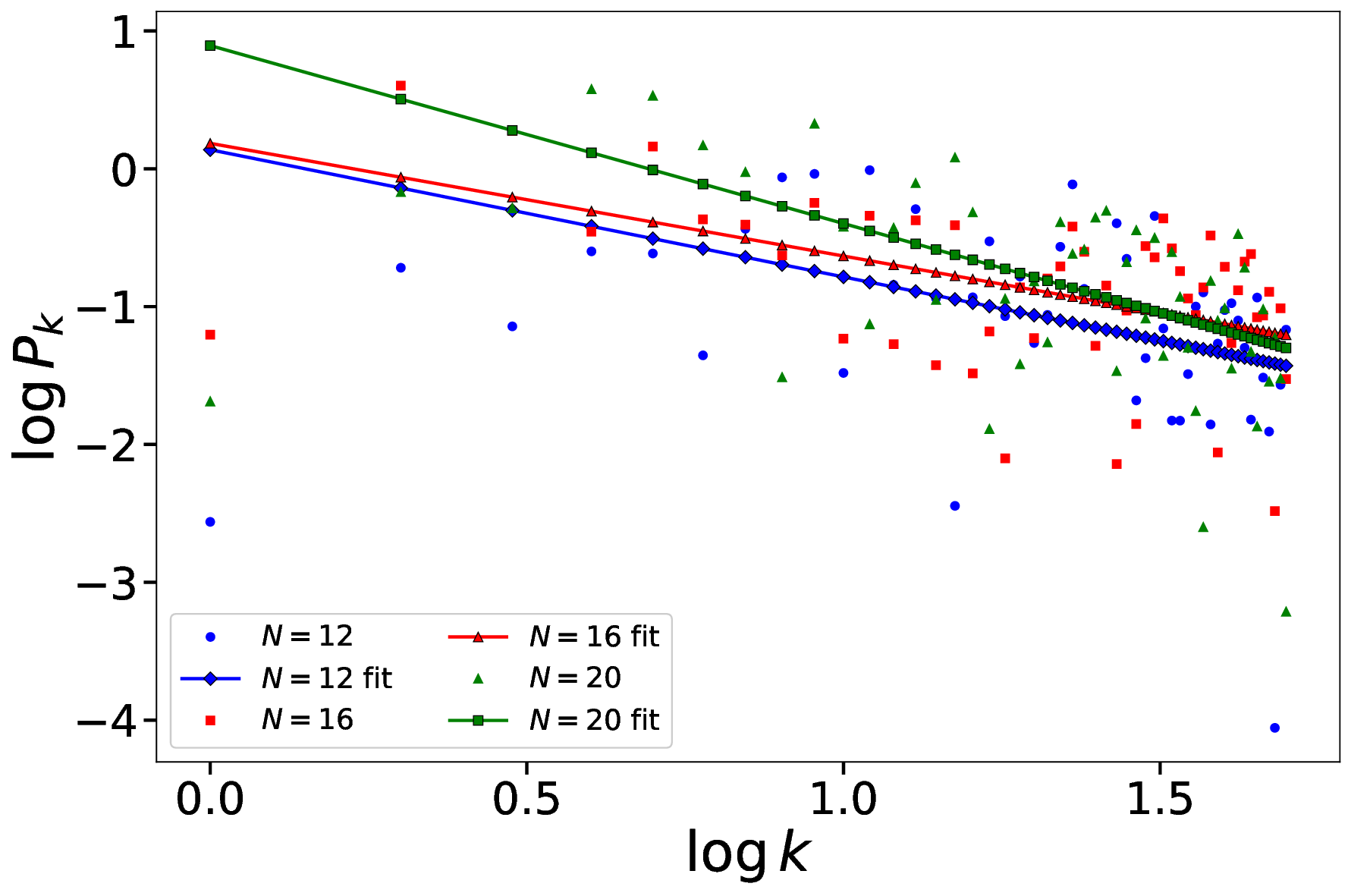}}
    \subfigure[$L_{z}=2N$]
   {\label{fig:ps5}\includegraphics[width=0.48\linewidth]{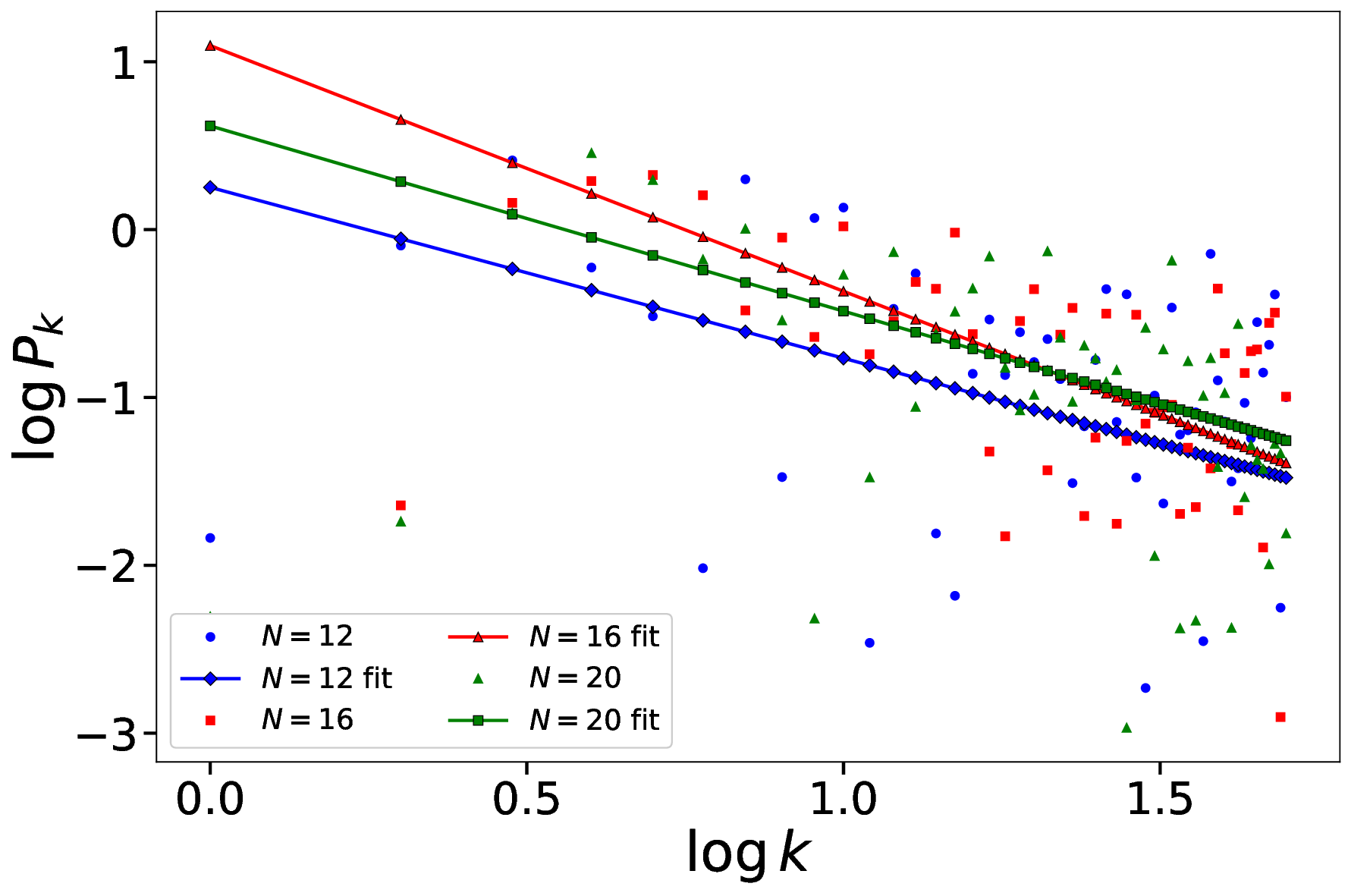}} 
    \subfigure[$L_{z}=3N$]
    {\label{fig:ps6}\includegraphics[width=0.48\linewidth]{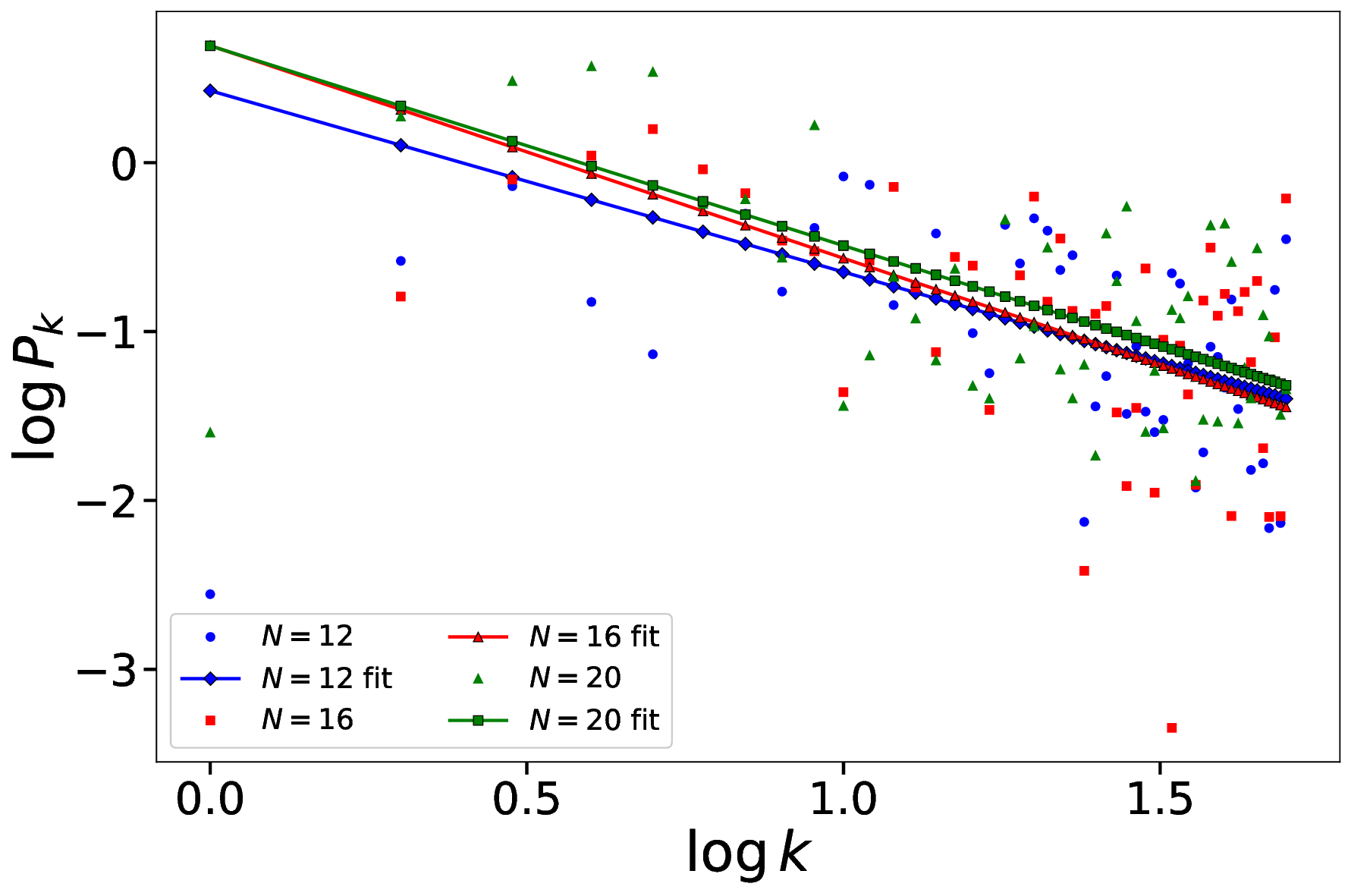}}

     \caption{The strong interaction regime: Power spectrum $P_{k}$ vs $k$ on the decimal log-log scale in the non-rotating (upper left), the single-vortex state (upper right), the multi-vortex states  $L_{z}=2N$ (lower left) and $L_{z}=3N$ (lower right) for $N=12, 16$ and $20$ bosons. The blue diamond, the red triangle and the green square lines are the straight line fits for our numerical data, denoted by dots, for $N=12, 16$ and $20$ bosons, respectively.}
    \label{fig:ps2}
\end{figure*}

\subsection*{a. Moderate interaction regime}The spectral form factor (SFF) for $N=12,16,$ and $20$ bosons for the non-rotating case, shown in the upper panel of Fig.~\ref{fig:sff1}, exhibits an initial oscillatory decay at early times, followed by suppression of oscillations and appearance of a minimum at dip time. For times longer than the Heisenberg time, the SFF saturates to a constant long-time value, giving rise to a plateau. Notably, the characteristic linear ramp expected in chaotic systems is absent, indicating the lack of significant spectral correlations and level repulsion predicted by RMT. This behavior of SFF signifies that the system remains in integrable regime, with weakly correlated energy levels, characteristic of Poissonian statistics. The integrability originates from the macroscopic occupation of a single-particle state due to Bose-Einstein condensation that leaves other states sparsely populated suppressing spectral correlations and resulting in SFF saturation without linear ramp. This is consistent with results on power spectrum $P(k)$, shown in the left panel of Fig.~\ref{fig:ps1}, which yields a power-law exponent of $\alpha \approx 2$, confirming Poissonian statistics. For the single-vortex state, the SFF exhibits an initial oscillatory decay, followed by a discernible linear ramp emerging after the dip time and eventually saturating to a plateau at long-time, as shown in the lower panel of Fig.~\ref{fig:sff1}. This behavior reflects vortex-induced partial level repulsion in the system. The formation of the single-vortex state gives rise to weak spectral correlations driving the system into pseudo-integrable regime, characterized by level statistics that deviates from integrability but does not exhibit universal correlations predicted by RMT. As the number of bosons increases from $N=12$, $16$ to $20$, the span of the linear ramp progressively decreases, as shown in Figs.~\ref{fig:12sff3}–\ref{fig:20sff3}, signaling reduction in spectral correlations and consequent tendency towards pseudo-integrability. This interpretation is further supported by the power spectrum analysis shown in the right panel of Fig.~\ref{fig:ps1}, where the extracted power-law exponents lie in the range $1<\alpha<2$ with values shown in Table \ref{powertable}.

\begin{table*}[htbp]
\caption{Values of power-law exponent $\alpha$ in the power spectrum with estimated error bar, $\Delta \alpha$: In the moderate interaction regime for the non-rotating and the single-vortex state; in the strong interaction regime for the non-rotating, the single-vortex and the multi-vortex states $L_z=2N, 3N$.}
\label{powertable}
\centering
\begin{ruledtabular}
\begin{tabular}{c c c c c c c}
$N$ &
\multicolumn{2}{c}{Moderate} &
\multicolumn{4}{c}{Strong} \\
%\multicolumn{2}{c}{Multi-vortex (strong)} \\
\cline{2-3} \cline{4-7} %\cline{6-7}
& \makecell{Non-rotating\\$\alpha \pm \Delta\alpha$} &  \makecell{Single-vortex\\$\alpha \pm \Delta\alpha$} & \makecell{Non-rotating\\$\alpha \pm \Delta\alpha$} & \makecell{Single-vortex\\$\alpha \pm \Delta\alpha$} & \makecell{$L_z=2N$\\$\alpha \pm \Delta\alpha$} & \makecell{$L_z=3N$\\$\alpha \pm \Delta\alpha$} \\
\hline
12 & 2.06 $\pm 0.238$  & 1.60 $\pm 0.354$ & 0.84 $\pm 0.349$ & 0.92 $\pm 0.353$ &  1.01 $\pm 0.706$ & 1.07 $\pm 0.220$ \\
16 & 1.73 $\pm 0.294$ & 1.57 $\pm 0.401$ & 1.58 $\pm 0.402$ & 0.82 $\pm 0.275$ &  1.46 $\pm 0.381$ & 1.25 $\pm 0.286$ \\
20 & 1.79 $\pm 0.279$ & 1.65 $\pm 0.356$ & 1.18 $\pm 0.256$ & 1.29 $\pm 0.261$ & 1.10 $\pm 0.518$ & 1.18 $\pm 0.224$ \\
\end{tabular}
\end{ruledtabular}
\end{table*}

\subsection*{b. Strong interaction regime}For the non-rotating case shown in the first row of Fig.~\ref{fig:sff2}, the SFF develops the characteristic dip-ramp-plateau structure. After reaching the dip time following the initial oscillatory decay, a small linear ramp emerges before saturating to a plateau at long times. The appearance of the distinct linear ramp signals the emergence of spectral correlations that is not strong enough to drive the system into GOE regime, indicating the system has moved further towards chaotic regime but continues to be in pseudo-integrable regime. The span of the linear ramp diminishes progressively with increasing number of bosons from $N=12$, $16$ to $20$, as shown in Figs.~\ref{fig:12sff2}, \ref{fig:16sff2} to \ref{fig:20sff2} and Table \ref{sfframptable}, indicating corresponding reduction in spectral correlations. The physical origin of pseudo-integrable behavior can be traced to interaction-induced depletion of the Bose-Einstein condensate, whereby bosons are scattered out of the macroscopically occupied phase-coherent single-particle state to other single-particle states with incoherent phases. This is further supported by the power spectrum results shown in the upper left panel of Fig.~\ref{fig:ps2}, where the extracted power-law exponents lie in the range $1<\alpha<2$, as shown in Table \ref{powertable}, indicating that the system resides in pseudo-integrable regime. For the single-vortex state shown in the second row of Fig.~\ref{fig:sff2}, the SFF develops a more pronounced linear ramp, evident from values of the span given in the fifth column of Table \ref{sfframptable}. This enhancement arises from two complementary mechanisms: strong interparticle interaction scatters bosons out of the macroscopically occupied single-particle state carrying one unit of angular momentum---the phase coherent Bose-Einstein condensate---to other single-particle angular momentum states with incoherent phases. The nucleation of the single-vortex state further facilitates transfer of bosons from the condensate to other sparsely occupied single-particle states, amplifying the depletion of the condensate. The combined effect of the above two mechanisms significantly enhances spectral correlations, leading to breakdown of integrability and driving the system to strongly chaotic regime consistent with RMT predictions. Accordingly, the power spectrum exhibits behavior close to GOE statistics, with fitted power-law exponents $\alpha \approx 1$ (values given in Table \ref{powertable} for $N=12, 16$ and $20$ bosons) as shown in the upper right panel of Fig.~\ref{fig:ps2}, confirming the emergence of chaos in the system. The SFF for the multi-vortex states with $L_{z}=2N$ and $L_{z}=3N$ is shown in the third and the fourth row of Fig.~\ref{fig:sff2}, respectively. Owing to the combined effects of the strong interparticle interaction and nucleation of quantized multi-vortices, the SFF exhibits a significantly enhanced linear ramp. This pronounced dip-ramp-plateau structure with enhanced linear ramp is a clear signature of strong quantum chaos and reflects substantial level repulsion, implying spectral rigidity in the underlying energy spectrum. The spans of the linear ramp measuring the degree of chaos in the system are summarized in Table \ref{sfframptable}. Physically, strong interaction scatters bosons out of the phase-coherent macroscopically occupied single-particle angular momentum state (the condensate) to other angular momentum states with incoherent phases, while rotation-induced nucleation of multi-vortices further transfers bosons out of the macroscopically occupied single-particle angular momentum states carrying two, three, or more units of angular momentum (depending on the vorticity) to sparsely occupied single-particle angular momentum states with incoherent phases. 
%The combined effect of these two mechanisms significantly enhances spectral correlations, driving the system into strongly chaotic regime. 
As the number of bosons increases from $N=12,16,$ and $20$, the linear ramp persists in the system, indicating continued presence of spectral correlations, as shown in the third and the fourth rows of Fig.~\ref{fig:sff2}. This is further corroborated by the power spectrum analysis, where the fitted lines for $N=12,16,$ and $20$ bosons closely follow GOE statistics with power-law exponent $\alpha \approx 1$, as shown in the lower left and the right panel of Fig.~\ref{fig:ps2} and summarized in Table \ref{powertable}, confirming the emergence of strong chaos in the system. It is noted from Table \ref{powertable} that the fitted power-law exponent in the power spectrum exhibits non-monotonic variations with number of bosons $N$, more specifically, for the non-rotating ($L_z=0$) and $L_z=2N$ states, in the strong interaction regime. Such variations of $\alpha$ with respect to $N$ are likely influenced by finite-size effects or the artifacts of small size of 100 levels of the energy-spectrum \cite{Relano2002}. For example, for the strongly interacting non-rotating case, the fitted exponents of power spectrum are $0.84 \pm 0.349$, $1.58 \pm 0.402$, and $1.18 \pm 0.256$ for $N=12$, $16$, and $20$, respectively. While the values of exponent $\alpha$ exhibit non-monotonic variations, the estimated error bars $\Delta \alpha$ indicate that the exponents are broadly consistent with the crossover from GOE ($\alpha \approx 1$) to Poisson statistics ($\alpha \approx 2$) as the number of bosons increases from $N=12$ to $16$ and $20$. Further, in the strongly interacting regime for $L_z=2N$ state, the fitted power-law exponents are $1.01 \pm 0.706$, $1.46 \pm 0.381$, and $1.10 \pm 0.518$ for
$N=12$, $16$, and $20$, respectively. The comparatively large values of error bars in the exponents may be attributed to the metastable \cite{hamid2022} nature of the $L_z=2N$ state. In contrast, the $L_z=3N$ state exhibits smaller error bars due to the stable nature of the state. The above results on power spectrum are consistent with the finite-size scaling analysis of the SFF.

\section{\label{4}Conclusion}In summary, we employ spectral form factor (SFF) and power spectrum as statistical tools to investigate crossover from integrability to quantum chaos in harmonically trapped interacting bosons in moderate to strong interaction regime, using exact diagonalization. We consider non-rotating and rotating Bose-Einstein condensate with single- and multi-vortex states. In the moderate interaction regime, the absence of linear ramp in SFF for the non-rotating case indicates that the system lies in integrable regime, whereas the appearance of a discernible linear ramp for the single-vortex state signals the emergence of pseudo-integrable behavior. The power spectrum yields the exponent $\alpha \approx 2$ for the non-rotating case and lies in the interval $1<\alpha <2$ for the single-vortex state corresponding to integrable and pseudo-integrable behavior, respectively, corroborating the results on SFF. In the strong interaction regime, the SFF displays a linear ramp of small span for the non-rotating case, indicating clear tendency towards chaotic regime but remains in pseudo-integrable region. As the system is subjected to external rotation leading to nucleation of single-vortex and multi-vortex states, the span of the linear ramp increases progressively, indicating the system now resides in strongly chaotic regime. The power spectrum yields the exponent in the interval $1<\alpha <2$ for the non-rotating case and $\alpha \approx 1$ for the single-vortex and multi-vortex states corresponding to pseudo-integrable and chaotic behavior, respectively, consistent with SFF scenario. As the number of bosons increases, the span of the linear ramp in SFF diminishes for the single-vortex state in the moderate interaction regime as also for the non-rotating case in the strong interaction regime, indicating weakening of spectral correlations and a tendency towards integrable behavior. In contrast, for the single- and multi-vortex states in the strong interaction regime, the linear ramp persists with increasing number of bosons, implying sustained presence of strong spectral correlations leading to chaotic behavior. It is anticipated that the spectral correlations in strong interaction regime for the single- and multi-vortex states will survive in the thermodynamic limit. Our study thus demonstrates that the crossover from integrable to chaotic behavior is intrinsically linked to depletion in the Bose-Einstein condensate---driven by interparticle interaction and nucleation of vortices. 
 
An extension of the present  study is the investigation of wavefunction-based measures, including the inverse participation ratio, information (Shannon) entropy and von Neumann entanglement entropy. These quantities probe eigenstate localization-delocalization and quantum correlations, thereby providing a complementary perspective to the spectral statistics. Some of these aspects of the study has been presented in our future communication \cite{talibipr}.

The interacting many-body bosonic system studied here does not possess a classical counterpart, however, an effective classical description can be obtained through mean-field approximation \cite{Castiglione_1996, Jona1996} using Gross-Pitaevskii equation \cite{dalfovo1999}. An interesting future work could be to establish direct connection between the spectral signatures of quantum chaos and the corresponding mean-field description of the system. In this framework, the degree of chaoticity can be quantified through Lyapunov exponent. Such an investigation forms a separate study and will be part of future work.

\begin{acknowledgments}
M.T. thanks the Ministry of Social Justice and Empowerment, Government of India, for Senior Research Fellowship (SRF).
\end{acknowledgments}

\end{document}